\documentclass[twocolumn]{aastex61}
\usepackage{amsmath, amstext}
\usepackage[T1]{fontenc}
\usepackage{apjfonts}
\usepackage[figure, figure*]{hypcap}

\def\lsim{\mathrel{\raise.3ex\hbox{$<$\kern-.75em\lower1ex\hbox{$\sim$}}}}
\def\gsim{\mathrel{\raise.3ex\hbox{$>$\kern-.75em\lower1ex\hbox{$\sim$}}}}
\def\gtwid{\mathrel{\raise.3ex\hbox{$>$\kern-.75em\lower1ex\hbox{$\sim$}}}}
\def\proptwid{\mathrel{\raise.3ex\hbox{$\propto$\kern-.75em\lower1ex\hbox{$\sim$}}}}

\newcommand{\OO}[1]{}

\shortauthors{Palumbo et al.}

\begin{document}



\title{Metrics and Motivations for Earth-Space VLBI: Time-Resolving Sgr A* with the Event Horizon Telescope}

\author{Daniel C. M. Palumbo}
\affil{Center for Astrophysics $|$ Harvard \& Smithsonian, 60 Garden Street, Cambridge, MA 02138, USA}

\author{Sheperd S. Doeleman}
\affil{Center for Astrophysics $|$ Harvard \& Smithsonian, 60 Garden Street, Cambridge, MA 02138, USA}

\author{Michael D. Johnson}
\affil{Center for Astrophysics $|$ Harvard \& Smithsonian, 60 Garden Street, Cambridge, MA 02138, USA}

\author{Katherine L. Bouman}
\affil{Center for Astrophysics $|$ Harvard \& Smithsonian, 60 Garden Street, Cambridge, MA 02138, USA}

\author{Andrew A. Chael}
\affil{Center for Astrophysics $|$ Harvard \& Smithsonian, 60 Garden Street, Cambridge, MA 02138, USA}

\begin{abstract}

Very-long-baseline interferometry (VLBI) at frequencies above 230\,GHz with Earth-diameter baselines gives spatial resolution finer than the ${\sim}50 \mu$as ``shadow'' of the supermassive black hole at the Galactic Center, Sagittarius A* (Sgr~A*). Imaging static and dynamical structure near the ``shadow'' provides a test of general relativity and may allow measurement of black hole parameters. However, traditional Earth-rotation synthesis is inapplicable for sources (such as Sgr~A*) with intra-day variability. Expansions of ground-based arrays to include space-VLBI stations may enable imaging capability on time scales comparable to the prograde innermost stable circular orbit (ISCO) of Sgr A*, which is predicted to be 4-30 minutes, depending on black hole spin. We examine the basic requirements for space-VLBI, and we develop tools for simulating observations with orbiting stations. We also develop a metric to quantify the imaging capabilities of an array irrespective of detailed image morphology or reconstruction method. We validate this metric on example reconstructions of simulations of Sgr~A* at 230 and\,345 GHz, and use these results to motivate expanding the Event Horizon Telescope (EHT) to include small dishes in Low Earth Orbit (LEO). We demonstrate that high-sensitivity sites such as the Atacama Large Millimeter/Submillimeter Array (ALMA) make it viable to add small orbiters to existing ground arrays, as space-ALMA baselines would have sensitivity comparable to ground-based non-ALMA baselines. We show that LEO-enhanced arrays sample half of the diffraction-limited Fourier plane of Sgr~A* in less than 30 minutes, enabling reconstructions of near-horizon structure with normalized root-mean-square error $\lesssim0.3$ on sub-ISCO timescales.

\end{abstract}


\section{Introduction} \label{sec:intro}
 A black hole leaves a dark imprint (the ``shadow'')  on nearby emission with a boundary shape dependent on black hole parameters \citep{Bardeen_1972,Falcke_2000}. An image of the bright accreting material near the event horizon provides an electromagnetic view of the local spacetime. Measuring the shadow size when the black hole mass is known (e.g., by studying stellar orbits as in \citealt{Ghez_2008}) provides a null hypothesis test of general relativity \citep{Psaltis_2015}. However, the dynamics of the matter surrounding the event horizon provide a more direct probe of parameters such as the black hole spin, which are difficult to extract solely from the shadow geometry \citep{Johannsen_2010}. For instance, the innermost stable circular orbit, or ISCO, is highly dependent upon spin, and can be studied by resolving periodicity near the event horizon \citep{Doeleman_2009, Fish_2009}.

Very-long-baseline interferometry (VLBI) enables angular resolution of the immediate vicinity of the largest known black holes. The Event Horizon Telescope (EHT) aims to image the immediate vicinity of the supermassive black holes in Sagittarius A* (Sgr~A*) and Messier 87 (M87) using a global network of radio telescopes which together provide high angular resolution through VLBI \citep{Doeleman_2009_decadal}. The 2018 configuration of the EHT observed at 230\,GHz, providing an effective angular resolution on Sgr~A* of $23\,\mu$as. This resolution is below the expected angular sizes of the black hole shadows in both Sgr~A* and M87. The mass to distance ratio is well-known for Sgr~A* and yields an expected shadow size of ${\sim}\,50\,\mu{\rm as}$ \citep{Gravity_2018}. This ratio is not as well known for M87, as gas and stellar dynamical results provide different mass estimates with corresponding shadow sizes of either ${\sim}\,20$ or ${\sim}\,40\,\mu{\rm as}$ \citep{Gebhardt_2011, Walsh_2013}. 

The combination of the EHT array and VLBI imaging algorithms designed to address the EHT's particular challenges is expected to be capable  of reconstructing static images of Sgr~A* at this resolution, and has done so for M87 \citep[see, e.g.,][]{Honma_2014, Bouman_2016,Chael_2016, Johnson_2017, Akiyama_2017a,Akiyama_2017b, Bouman_2018,Kuramochi_2018,Chael_closure,PaperIV}. However, imaging time-variable structure around supermassive black holes requires well-sampled spatial baseline coverage (conventionally described in the $(u,v)$ plane) on timescales comparable to the innermost stable circular orbit (or ISCO). Though the current EHT provides sufficient angular resolution to image the shadow of both Sgr~A* and M87, the array does not provide sufficient instantaneous (or ``snapshot'') coverage to reconstruct a rapidly time-varying source intensity distribution at Sgr~A*, as we explore later. The $(u,v)$ sampling of ground-based arrays is fundamentally limited by the speed of Earth rotation; thus, many sites are required to attain comprehensive ``snapshot'' coverage of rapidly evolving sources.

The EHT plans to observe at 345\,GHz in the near future. This higher frequency will provide several advantages when observing Sgr~A*: the magnitude of interstellar scattering effects decreases with the square of the observing wavelength $\lambda$, and the diffraction-limited angular resolution ($\lambda/D$) improves \citep[see, e.g.,][see also \citealt{Johnson_2016,Johnson_Narayan_2016,Psaltis_2018_scattering}]{Harris_1970,Narayan_1992}. However, observing at 345\,GHz also introduces new challenges: receiver sensitivity decreases due to higher system temperature and atmospheric phase fluctuations increase, thereby limiting the feasible coherent integration time of VLBI observations before calibration \citep{TMS}. Furthermore, dishes require higher surface accuracy at high frequencies in accordance with Ruze's Law, favoring smaller dishes that more easily meet these specifications \citep{Ruze_1966}. 

In this paper, we develop a methodology for analyzing space-VLBI arrays. We then explore a possible future development of the EHT: expanding the array to include dishes in Low Earth Orbit (LEO), enabling time-domain analysis and dynamical imaging reconstructions of Sgr~A*. Space dishes in low-Earth orbit provide benefits to imaging due to the rapid formation of baselines to ground dishes with many different lengths and orientations. To match the next generation EHT, we generally use 345\,GHz as the simulated frequency of observation for our analysis, though the differences in imaging at 230 and 345\,GHz are discussed. In \autoref{sec:background}, we review prior work on Sgr~A* with VLBI, and we examine theoretical constraints and prior space-VLBI missions to inform our investigation of a LEO expansion to the EHT. 
In \autoref{sec:fourier}, we develop a pre-imaging metric for array performance, and we demonstrate the value of adding space dishes for improving the angular and temporal resolution of the EHT. 
In \autoref{sec:imaging}, we compare examples of static and dynamical reconstructions of simulated models observed with ground and space-enabled arrays. We apply simple image-domain feature extraction algorithms to reconstructions of a general relativistic magnetohydrodynamic (GRMHD) simulation of Sgr~A* and demonstrate the necessity for algorithmic development focused on temporal observables in the image domain.
In Section~\ref{sec:discussion}, we briefly discuss the parameter space of sensitivity that may inform a future hardware study, and look to other concepts for space-VLBI as well as areas in need of further examination.

\section{Background}
\label{sec:background}

Though the EHT is already nominally capable of reconstructing images of static structure at Sgr~A*, the array likely requires expansion to image the time-varying structure that is expected to exist at the event horizon scale. Small space dishes may efficiently address this requirement, but geometrical restrictions on orbiting VLBI dish performance present challenges that we now consider in detail. Here we present the considerations of source evolution, existing ground stations, past space-VLBI missions, and analytic constraints that motivate and inform a time-domain focused expansion of the EHT to space. 

\subsection{Sagittarius A*}

Sgr A*, the radio source at the center of our galaxy, is coincident with a $4.1 {\times}10^6\,\textrm{ M}_\odot$ black hole at a distance of 8.1\,kpc from Earth \citep{Ghez_2008,Gravity_2018}. Sgr~A* is expected to have a shadow that subtends ${\sim}\,50\,\mu{\rm as}$, making it the largest known black hole as seen from Earth. In order to resolve the shadow, an observing instrument must have a diffraction-limited resolution finer than the shadow size.

Meanwhile, properties of the emission from Sgr A* limit the observing frequency. Observations and theoretical predictions of Sgr A* indicate synchrotron radiation in near-horizon emission \citep[see, e.g.,][]{Yuan_2003,Bower_2015,Chael_2018}. At long wavelengths, the local plasma is optically thick to synchrotron radiation, leading to synchrotron self-absorption that obscures event-horizon scale structure \citep[see, e.g.][]{Blandford_1999,Chan_2015, Davelaar_2018}. Thus, observations of the black hole shadow must occur at higher radio frequencies at which the accretion flow is optically thin.

Radio emission from the galactic center scatters predominantly off of cold plasma in the ionized interstellar medium with a dispersion relation that depends on the local electron density \citep{Kulsrud_2005}. Perturbations to the electron density cause delays in the phase velocity of an emitted signal, leading to warped radio images \citep{Johnson_Narayan_2016}. The characteristic angle of the associated refractive effect scales with the square of the observing wavelength; at high radio frequencies, there persists a small but non-negligible diffractive blurring effect with refractive substructure. Though the blurring angle is smaller than the nominal 230 GHz beam of the EHT, tools have been developed to mitigate this fundamental limit on VLBI images of objects in the Galactic plane \citep{Doeleman_2009_decadal,Johnson_2016}. The 230 and 345 GHz observing bands considered by the EHT fall within windows of transparency for Earth's atmosphere, enabling observation from the ground.

Sgr~A* has been observed at many frequencies to be intensely time-varying on timescales as short as 30 minutes \citep[see, e.g., at mm/sub-mm:][see also Near-Infrared/X-Ray results in \citealt{Baganoff_2001,Genzel_2003,Aschenbach_2004,Ghez_2004, Belanger_2006,Meyer_2006,YZ_2006, Hornstein_2007,Dodds-Eden_2011,Neilsen_2013,Ponti_2017,Gravity_2018}]{Miyazaki_2004,YZ_2006,Marrone_2008,Bower_2015}.
The rapid time variability of Sgr~A* provides both challenges for imaging and opportunities for science 
beyond improving reconstruction of the black hole shadow \citep{Lu_2016}. The size and shape of the shadow is weakly dependent on spin, yet the ISCO period of Sgr~A* varies between 4 minutes for a maximal spin black hole and half an hour for a black hole with zero spin \citep{Bardeen_1973,Takahashi_2004,Johannsen_2010}. If there is variation in the source intensity distribution on timescales similar to the ISCO period, an observing VLBI array would need well-sampled baseline coverage on ${\sim}\,30$ minute timescales in order to reconstruct instantaneous images of the source dynamics (though non-imaging time-domain methods may have different sampling requirements as in \citealt{Doeleman_2009,Fish_2009}). Further, recent near-infrared astrometric and polarization measurements of the Galactic Center suggest orbital motion on ISCO timescales that is likely visible in the angular region to which the EHT is sensitive \citep{Gravity_2018_orbit}. Temporally resolving Sgr A* with an expanded EHT thus provides an opportunity to connect measured variability from other frequency regimes with imaged source dynamics at the event horizon scale.

\subsection{The Event Horizon Telescope}\label{sec:EHT}

\begin{figure}[t]
\centering
\includegraphics[width=.45\textwidth]{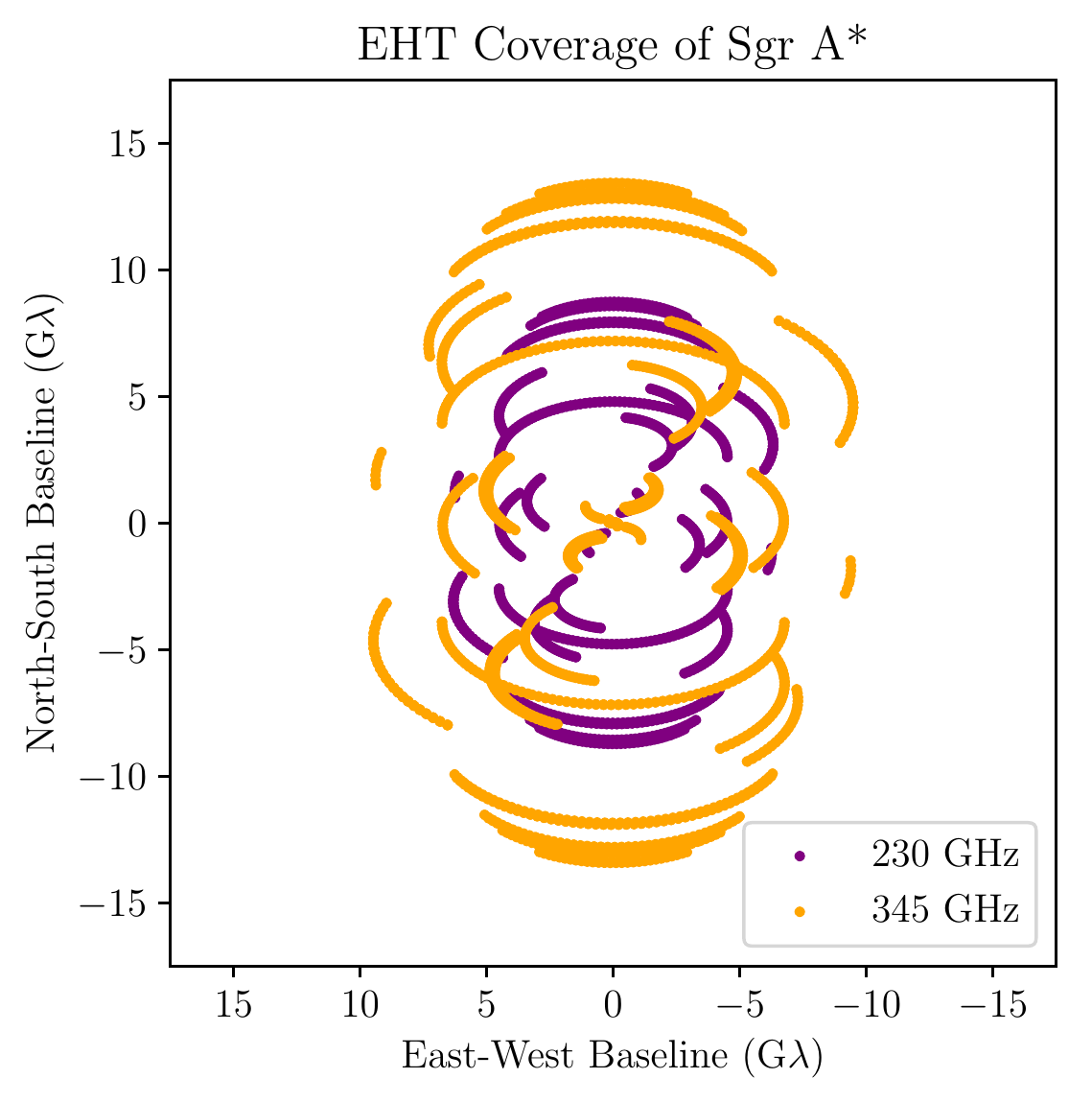}
\caption{Baseline coverage of Sgr~A* provided by the 2019 EHT at 230 GHz and the future 345 GHz EHT (or ``EHTII'') considered in this paper. The 2019 array includes PV, SMT, SMA, ALMA, SPT, APEX, JCMT, and LMT. The EHTII array is simulated with sites at Kitt Peak and in the French Alps.}
\label{fig:eht2018_uv}
\end{figure}

As of the April 2018 observing campaign, the EHT contains 8 telescopes that observe Sgr~A* from 6 geographic sites: the Atacama Large (sub)-Millimeter Array, or ALMA, in Chile; the Atacama Pathfinder Experiment Telescope, or APEX, also in Chile and very close to ALMA; the James Clark Maxwell Telescope, or JCMT, near the summit of Mauna Kea in Hawaii; the Large Millimeter Telescope, or LMT, in Mexico; the 30-meter telescope on Pico Veleta in Spain operated by the Institut de Radioastronomie Millim\'{e}trique, or PV; the Submillimeter Array, or SMA, located near the JCMT; the Submillimeter Telescope, or SMT, located on Mount Graham in Arizona; and finally, the South Pole Telescope, or SPT, operating at the National Science Foundation's South Pole research station. Two additional sites are expected to join the Event Horizon Telescope array in the near future: the Kitt Peak National Observatory, or KP, and the Northern Extended Millimeter Array, or NOEMA, in the French Alps. The EHT also includes the Greenland Telescope, though it can not observe Sgr~A*. The simulated observations in this article include these dishes with realistic hardware estimates to approximate the future EHT; hereafter we refer to this array as ``EHTII.''

EHT stations span a large range of antenna separations, running from ``trivially separated''  dishes with ${\sim}100$\,k$\lambda$ baselines to distant telescopes with ${\sim}13$\,G$\lambda$ baselines at 345\,GHz. In a full day of observation, the EHT array has sufficiently well-sampled baseline coverage of near-equatorial sources to form static images \citep[see, e.g.,][]{Chael_2016}, though coverage along the northeast-southwest direction is particularly sparse - see \autoref{fig:eht2018_uv} for the full-day $(u,v)$ coverage of the approximately $-29^\circ$ declination of Sgr A*.

\subsection{Basic Requirements for Space-VLBI}

VLBI baselines measure complex-valued spatial Fourier components (``visibilities'') of the source brightness on the sky by correlating co-temporal measurements of the electric field across large distances. As stations move in the orthographically projected plane of the Earth as seen from the source, different Fourier components are measured as ``tracks'' are swept in the $(u,v)$ plane, as in \autoref{fig:eht2018_uv}. These ``tracks'' are typically ellipses corresponding to the shift in the displacement vector between two ground-based sites; for space dishes, these tracks correspond to instantaneously elliptical paths with time-dependent semi-major axes.

Visibility measurements are corrupted by instrumental and atmospheric gain variations discussed in detail in \citet{TMS}. Orbiting VLBI stations face different observation parameter demands than ground-based stations. For example, the integration time $\tau$ is limited by the timescale of phase coherence. Neglecting reference hardware coherence, ground site phase coherence is dominated primarily by turbulence in the atmosphere. However, for orbiting VLBI stations, the dominant constraints on the integration time arise from thermal noise and the speed of the orbiter through the $(u,v)$ plane.

The motion of VLBI observing sites is crucial to Fourier synthesis, but also introduces fundamental limitations on integration time.\ As the baseline vector $\vec{u}$ rotates, the phase of the visibility measurement rotates, eventually picking up a full phase wrap over the course of one averaged measurement. \citet{TMS} provide a bounding condition on integration time to prevent a phase wrap, formalized for a source confined to within an angle $\theta_{\textrm{FOV}}$:
\begin{align}
\tau < \frac{1}{ \omega D_\lambda \theta_{\textrm{FOV}}}. 
\label{eq:tau_orbit}
\end{align}
Here, $\omega$ is the angular velocity of the rotation of the observing site, $D_\lambda = |\vec{u}_{\textrm{max}}|$ is the length of the longest baseline in wavelengths, and $\theta_{\textrm{FOV}}$ is in radians. For a nearly-circular Low Earth Orbit (as we examine later), the rotation rate is $\omega = \frac{2 \pi}{P}$ with $P \approx 1.5$ hours. We are interested primarily in filling in gaps in existing $(u,v)$ coverage, so we focus on coherent averaging measurements out to the maximum baseline of a LEO-enabled array, giving $D_\lambda \approx 15$\,G$\lambda$ at 345 GHz. We further assume that the source structure of interest is confined to a circular angular extent of diameter $180\,\mu{\rm as}$, sufficient to contain multiple shadow-scales, though likely not to image extended structure, such as a jet. These values together yield $\tau \lesssim 1$ minute, giving a bound on coherent averaging of 30 seconds \citep{TMS}.

To generalize the coherence time metric to satellites with arbitrary orbital semi-major axis $a_{\rm orb}$ and eccentricity $e$, we must find the maximum instantaneous angular velocity for an eccentric orbit. Conservation of mechanical energy yields the \textit{vis-viva} equation for the orbital speed $v_{\rm orb}$,
\begin{align}
    v_{\rm orb} = \sqrt{\mu \left(\frac{2}{r} - \frac{1}{a_{\rm orb}}\right)},
\end{align}
where $r$ is the instantaneous distance of a small mass from the Earth center of mass and $\mu=GM$ is the gravitational parameter, simplified to the product of the gravitational constant $G$ and the Earth mass $M$. The maximum instantaneous angular velocity occurs at periapsis, where $\omega_{\textrm{max}}$ is given by $v_{\rm orb}/r$ when $r = a_{\rm orb} (1-e)$, yielding:
\begin{align}
    \omega_{\textrm{max}} = \sqrt{\frac{\mu(1+e)}{a_{\rm orb}^3 (1-e)^3} }.
    \label{eq:bound}
\end{align}
Assuming that the integration time is held constant throughout the orbit requires that the bound (\autoref{eq:bound}) hold for the longest baseline in the orbital geometry, which occurs approximately at apsis; for an orbit with apsis inclined at an angle $\psi$ relative to the source line-of-sight (with $\psi = \pi/2$ corresponding to the ``face-on'' orbit described later), 
\begin{align}
    D_\lambda \approx \frac{a_{\rm orb} (1+e) \sin{\psi}}{\lambda}
\end{align}
neglecting motion of ground sites. This relation holds only if the longest baseline in the array is comparable to the baseline from the orbiter to the center of the Earth, as would be the case for a VLBI array with only one orbiter far from the Earth. Otherwise, in the case of an array with, e.g., two diametrically opposed orbiters, or one orbiter with $a_{\rm orb}$ comparable to the Earth radius (as is the case for the LEO orbits we consider), this approximation should be increased by a factor of 2 (denoted by brackets in the equation below). Substituting this approximation and our expression for $\omega_{\textrm{max}}$ into \autoref{eq:tau_orbit} gives
\begin{align}
    \tau_{\textrm{max}} \approx \frac{\lambda}{[2]\theta_{\textrm{FOV}} \sin{\psi}} \sqrt{\frac{a_{\rm orb} (1-e)^3}{\mu (1+e)^3}}.
\end{align}
For the LEOs discussed in \autoref{sec:fourier}, $e=0$ and $a_{\rm orb}$ is approximately equal to the Earth radius. Taking the factor of 2 into account and using $\theta_{\rm FOV} = 180 \mu{\rm as}$, $\lambda = 0.87$mm and $\psi = \pi/2$ recovers the $\tau_{\rm max} \lesssim 1$ minute found earlier. 

The sensitivity of an individual station is described by its system equivalent flux density, or SEFD, which is given in terms of the Boltzmann factor $k_{\rm B}$, the system temperature $T_{\rm sys}$, and the effective collecting area $A_{\rm eff}$:
\begin{align}
    {\rm SEFD} = \frac{2 k_{\rm B} T_{\rm sys}}{A_{\rm eff}}.
\end{align}
The sensitivity of a particular baseline is described by its thermal noise, which depends on the SEFDs of its constituent stations. The thermal noise is given by \citep{TMS}
\begin{align}
\sigma = \frac{1}{\eta_\textrm{Q}} \sqrt{\frac{\textrm{SEFD}_1 \text{SEFD}_2}{2 \Delta \nu \tau}},
\label{eq:sigma}
\end{align}
where $\Delta\nu$ is the observing bandwidth and $\eta_Q$ is a digital correction factor due to finite quantization of the received radio emission. If 2-bit quanitization is used (as in the current EHT), $\eta_{\rm Q}=0.88$.

Small dishes contribute effectively to VLBI when forming baselines to highly sensitive stations such as ALMA because the thermal noise depends on the geometric mean of the sensitivities of the constituent dishes. The LMT may also be suitable as an ``anchor'' station for small dishes, should it observe at 345 GHz. The recently coherently phased ALMA now has an SEFD at millimeter wavelengths on the order of ${\sim}\,100$ Jy \citep{Matthews_2018}. For the purposes of our small-dish sensitivity computations, we use an orbiter with a diameter of 4m. We assume an aperture surface efficiency of 80\%; other factors such as illumination, blockage, etc., can also contribute to the total aperture efficiency.

The ${\sim}4$m class of dish has been successfully launched in a non-deployable architecture \citep[see, e.g. the \textit{Herschel} instrument,][]{Pilbratt_2010}. Deployable architectures may also be suitable for high-frequency performance \citep{Millimetron,Datashvili_2014}. We note, however, that 4m is not an optimized diameter, and is adopted simply as a benchmark ``small dish'' for the example calculations and reconstructions that follow.

We thus compute the 345\,GHz SEFD of a 4m dish to be ${\sim}\,20000$\,Jy, where we estimate the atmosphere-free system temperature to be 75\,K at 345\,GHz (found by assuming similar performance to ALMA receivers at band 7 as in \citealt{Matthews_2018}). Using a ${\sim}150$ Jy estimated zenith SEFD of phased ALMA at 345 GHz, we can compute a minimum integration time $\tau_{\textrm{min}}$ based on a desired nominal thermal noise $\sigma_{\textrm{nom}}$ by rearranging \autoref{eq:sigma}:
\begin{align}
\tau_{\rm min} = \frac{\rm SEFD_1 SEFD_2}{2\Delta\nu}\Big(\frac{1}{\eta_\textrm{Q} \sigma_{\rm nom}}\Big)^2.
\end{align}
We choose a desired thermal noise of 10 mJy based on long-baseline $({\sim}\,\rm 7$~G$\lambda)$ correlated flux densities of tenths of Janskys observed for Sgr~A* \citep{Lu_2018}. This approximate mean sensitivity over a full observing track yields a required $\tau_{\textrm{min}} \approx 1$ second for space-ALMA baselines. Between the same LEO dish and a more typical ground site with SEFD $\approx 10000$\,Jy, $\tau_{\textrm{min}} \approx 80$ seconds. Space-ALMA baselines are thus necessary to reach ground-comparable signal quality within the motion-based decoherence of the VLBI signal. For the simulated observations presented in this article, we maintain the integration time at the 30 second limit from \autoref{eq:tau_orbit}, guaranteeing detections to ALMA without exceeding the motion-based limit. Space-ALMA detections would then allow calibration of all other space-ground baselines on timescales shorter than the 80 second thermal noise bound \citep[see, e.g.,][ for examples of network calibration with ALMA]{PaperIII}.

\subsection{Past Efforts in Orbiting VLBI}\label{sec:past}

The first Earth-space fringe detection was in 1986, using the Tracking and Data Relay Satellite System (or TDRSS) system in geostationary orbit at observing frequencies of 2.3 and 15\,GHz \citep{Levy_1989}. Non-geostationary orbits sweep through much broader baseline coverage and are not fundamentally limited in baseline length; in 1997, the VLBI Space Observatory Programme, or VSOP, brought the 8-m diameter Highly Advanced Laboratory for Communications and Astronomy (HALCA) into an elliptical Earth orbit with a period of approximately 6.6 hours and an apogee of 21,000 km \citep{Hirabayashi_2000}. HALCA was followed by the 10-m diameter RadioAstron (or Spektr-R) \citep{Kardashev_2013}, with a period of 8.6 days and an apogee of approximately 300,000 km. These missions operated at centimeter wavelengths and successfully detected fringes despite the difficulties of space-ground VLBI. Though some of these projects had a planned angular resolution similar to the EHT (see \autoref{tab:space_missions}), none was operating in the high-frequency regime required to overcome the interstellar scattering of emission from Sgr~A*, which obscures near-horizon structure at wavelengths as low as 3\,mm \citep{Issaoun_2019}. These projects provide partial guidance for future efforts in space-VLBI.

\begin{table*}[t]
\caption{Previous space-VLBI Missions}
\label{tab:space_missions}
\begin{tabular}{lllllllll}
\hline
\hline
Platform & Diam. & Observing Freq{.} & Bandwidth & Period & Apogee & Nom. Res. & Years Active & Reference\\
\hline
TDRSS & 4.9 m & 2.3, 15\,GHz & 14\,MHz & 24\,hr & 42164\,km & ~100\,$\mu$as & 1986-1987 & \cite{Levy_1989}\\
\hline
VSOP & 8 m & 1.6, 5, 22\,GHz & 32\,MHz & 6.3\,hr & 21400\,km & ~130\,$\mu$as& 1997-2005 & \cite{Hirabayashi_2000}\\
\hline
RadioAstron & 10\,m & 0.3, 1.6, 4.8, 22\,GHz & 32\,MHz & 8.5\,d & 371000\,km & ~7\,$\mu$as& 2011-present & \cite{Kardashev_2013}\\
\hline

\end{tabular}
\end{table*}

\section{Baseline Coverage}
\label{sec:fourier}

Orbiting VLBI elements are not bound by the surface or rotation rate of the Earth, and can thus form a broader range of baselines to stations on the ground on shorter timescales than afforded by Earth-rotation Fourier synthesis. In particular, the orientation and period of the orbit can be chosen to fill in gaps in the existing coverage of the array with greater flexibility than is possible for a ground site, for which one must account for such factors as altitude, weather, and infrastructural support. Here we present a simple example orbit for rapid filling of $(u,v)$ coverage of Sgr~A* with space-ground and space-space baselines when observing with the ``EHTII'' array. 

\subsection{Orbit Design and Simulation}

We consider expanding the EHT to space in order to improve instantaneous baseline coverage for dynamical imaging of Sgr~A*. This particular hypothetical space-enabled EHT differs from previous space-VLBI missions such as VSOP and Spektr-R (and its upcoming follow-up, Spektr-M) and from other possible EHT expansion paradigms in that we assume the ground-based EHT already provides sufficient angular resolution to resolve the black hole shadow of Sgr~A*, and do not pursue major improvements to angular resolution with longer space-ground baselines. 

Instead, we utilize orbiting components of the EHT array to fill in gaps in existing coverage over short timescales. In the current  EHT, large regions of missing $(u,v)$ coverage (\autoref{fig:eht2018_uv}) limit the fidelity and dynamic range of reconstructed images. Filling holes in the sampled $(u,v)$ plane reduces the magnitude of sidelobes in the Fourier transform of the synthesized visibility measurements (or ``dirty beam''), generically improving image reconstructions across algorithms.

To model space dishes operating in concert with the EHT, we developed software to manipulate Two-Line Element sets (or TLEs) and simulate VLBI observations with space dishes. This software creates synthetic TLEs for arbitrary orbital elements that are compatible with any Simplified Perturbation Model-based orbit calculator \citep[see, e.g.,][]{Wei_2010}. Further, we can time-delay existing TLEs to precisely shift orbital phase to any time relative to an EHT observing window, though we do not perform such an optimization in this study. Instead, we choose an observation time in Greenwich Mean Sidereal Time (or GMST) at which most EHT ground stations can see the source; for an observation longer than approximately half of an orbital period, most baselines of interest will be sampled, meaning that the initial phase is largely irrelevant. 

\begin{figure}[t]
    \centering
    \includegraphics[width=.45\textwidth]{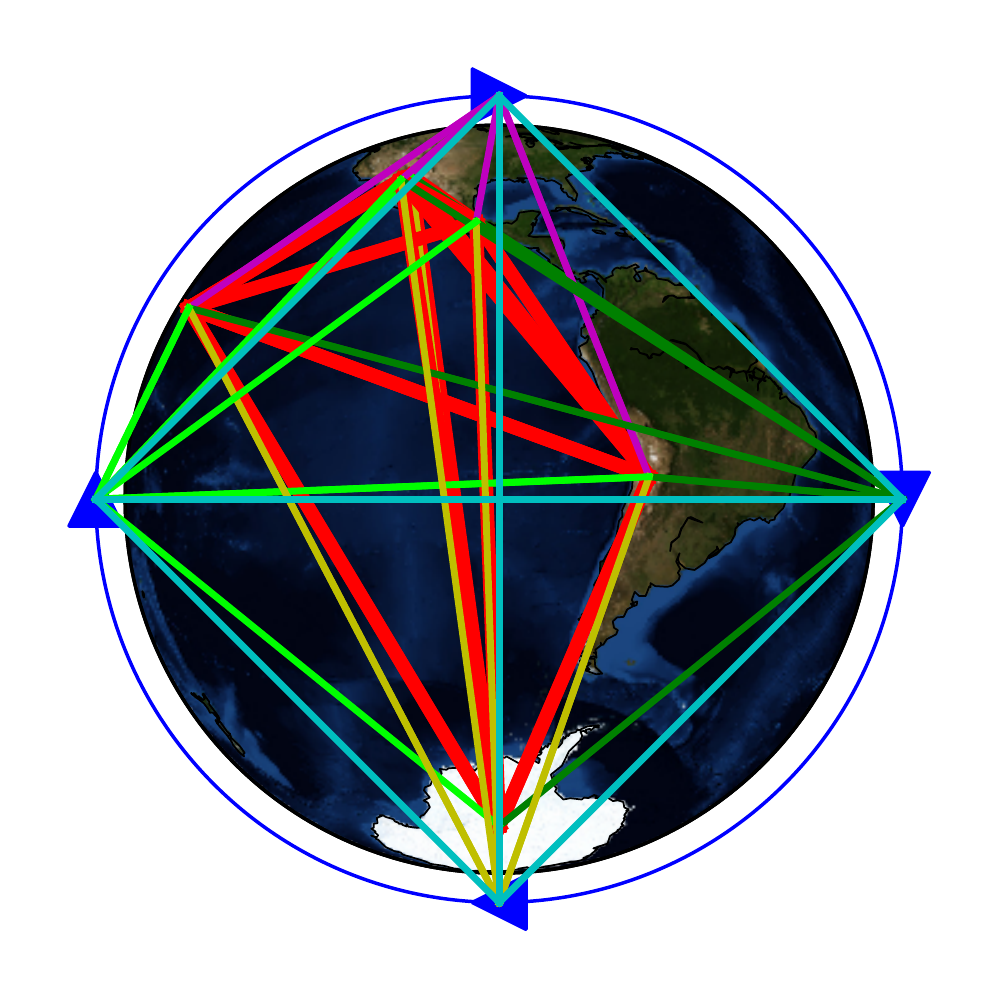}
    \caption{Schematic diagram of a possible expansion to the EHT, pictured as 4 Low Earth Orbiters always in view of Sgr~A*. Thick red baselines are shown at 0~GMST for the EHT's expected ground-based 345\,GHz configuration, ``EHTII.'' Blue arrows correspond to orbiter positions; cyan lines are space-space baselines. Other lines are space-ground baselines grouped by color for each orbiter. Over the course of a full 90 minute orbit, each orbiter contributes baselines across a wide range of $(u,v)$ separations.}
    \label{fig:orbit_diagram}
\end{figure}

\begin{figure}[t]
\centering
\includegraphics[width=.45\textwidth]{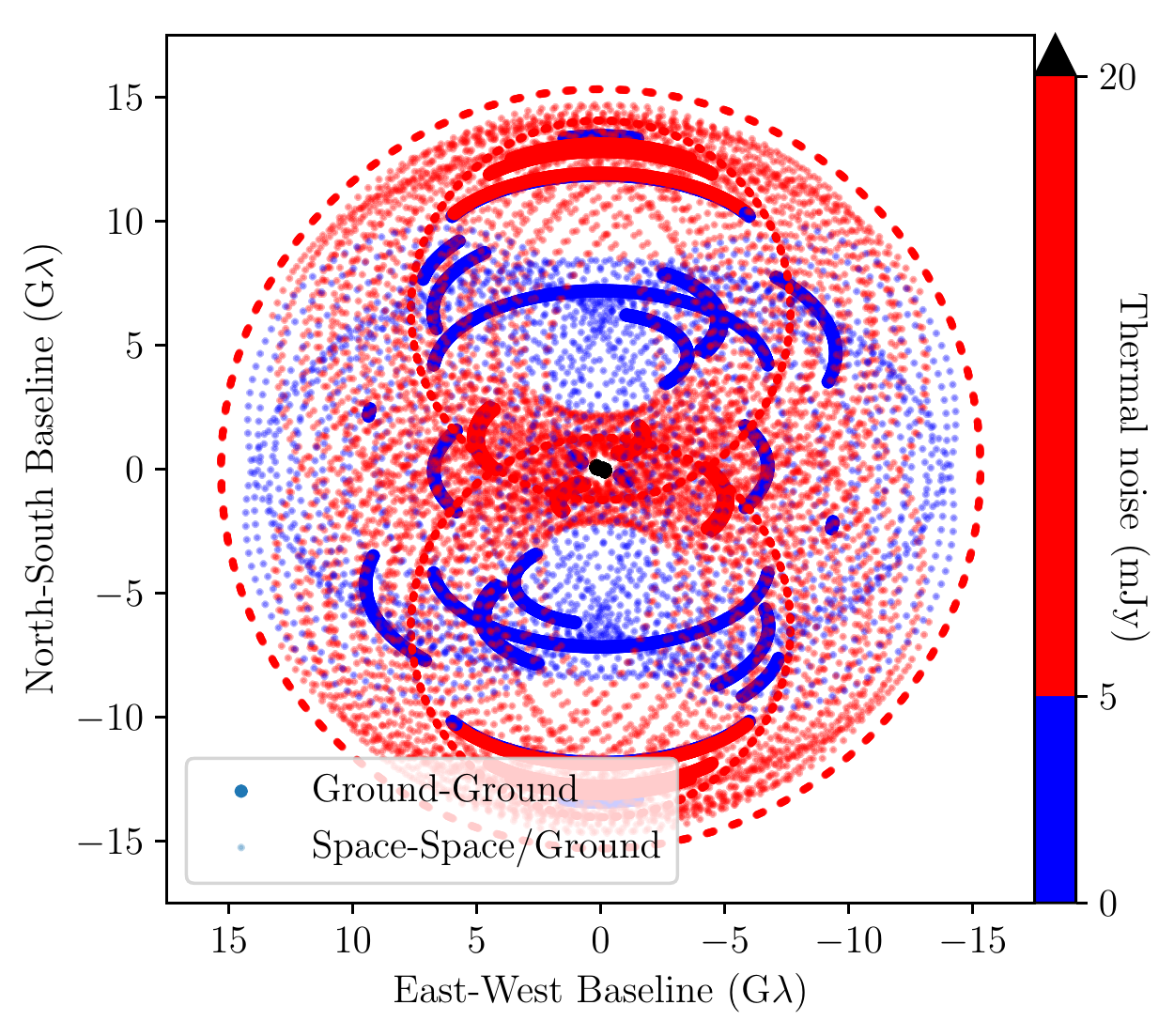}
\caption{Sgr~A* baseline coverage of two diametrically opposed face-on orbiters over 24 hours with SEFD $\approx 20000$ Jy observing with the EHT at 345\,GHz. Measurements are shown every 60\,seconds, and are colored by thermal noise as computed for a bandwidth of 16\,GHz, integration time of 30\,seconds, and zenith site opacities estimated from early EHT data. Measurements with less than 5~mJy of thermal noise are considered to be highly sensitive detections, and are shown in blue. Measurements with thermal noise between 5-20\,mJy are shown in red. Measurements with higher thermal noise are shown in black. Black points only include the PV-NOEMA baseline, which has high thermal noise due to the low elevation of Sgr~A*; as the baseline is short, these measurements would still have high S/R. Notably, baselines between space dishes are not impacted by atmospheric opacity, and thus have comparable thermal noise to ground sites with superior nominal sensitivity.}
\label{fig:1ring_cov}
\end{figure}

\begin{figure*}[t]
    \centering
    \includegraphics[width=.3402\linewidth]{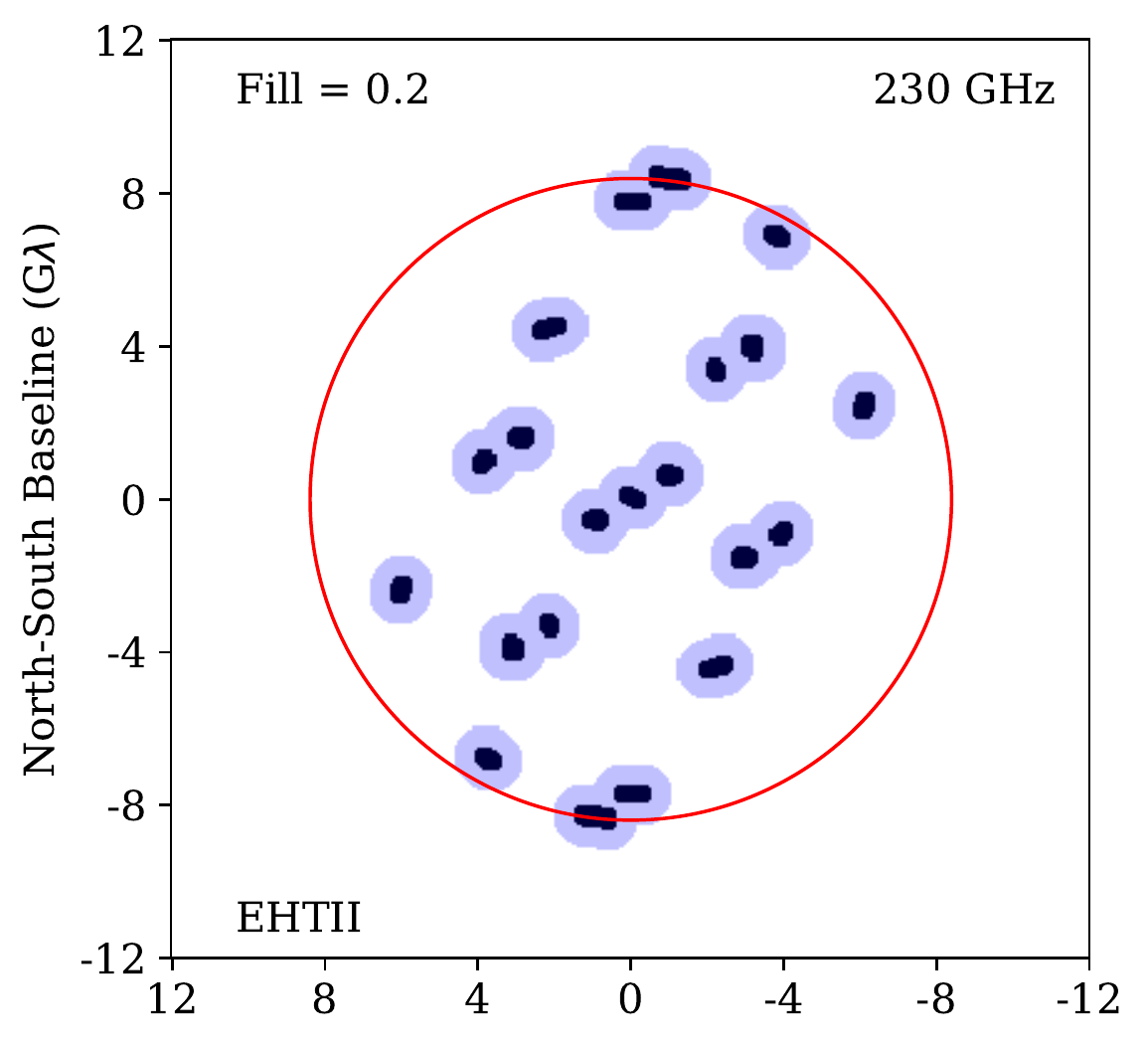}
    \includegraphics[width=.324\linewidth]{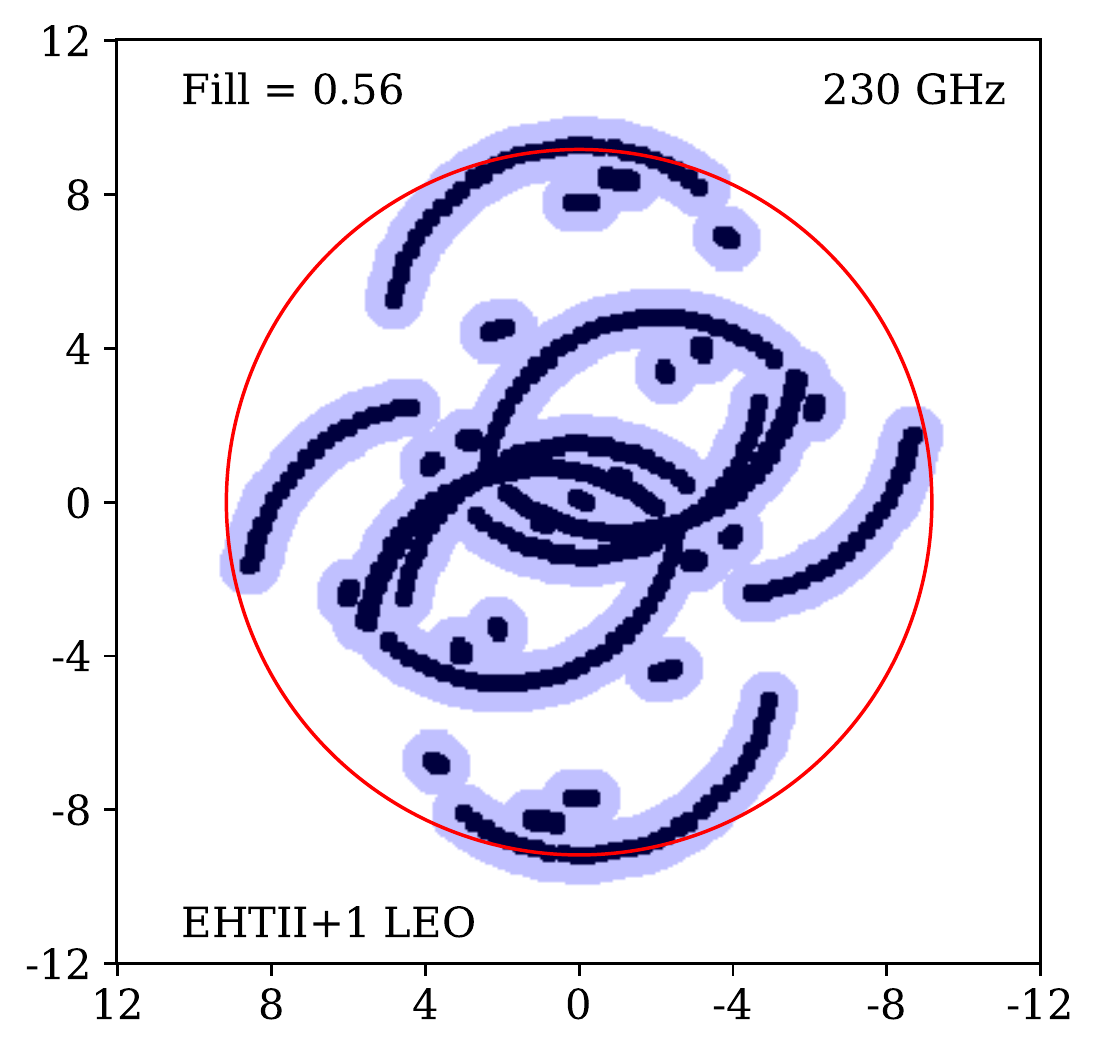}
    \includegraphics[width=.324\linewidth]{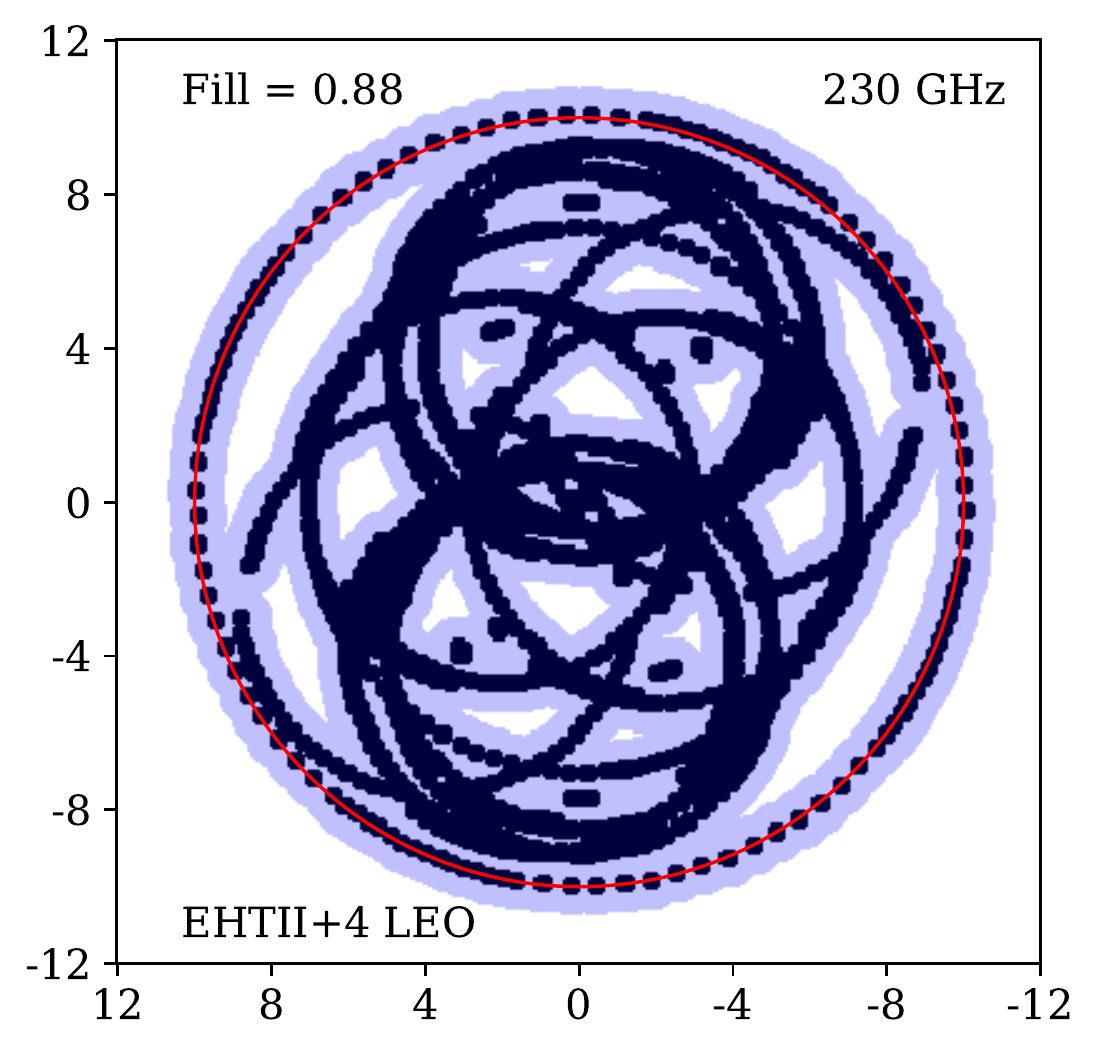}
    \includegraphics[width=.3402\linewidth]{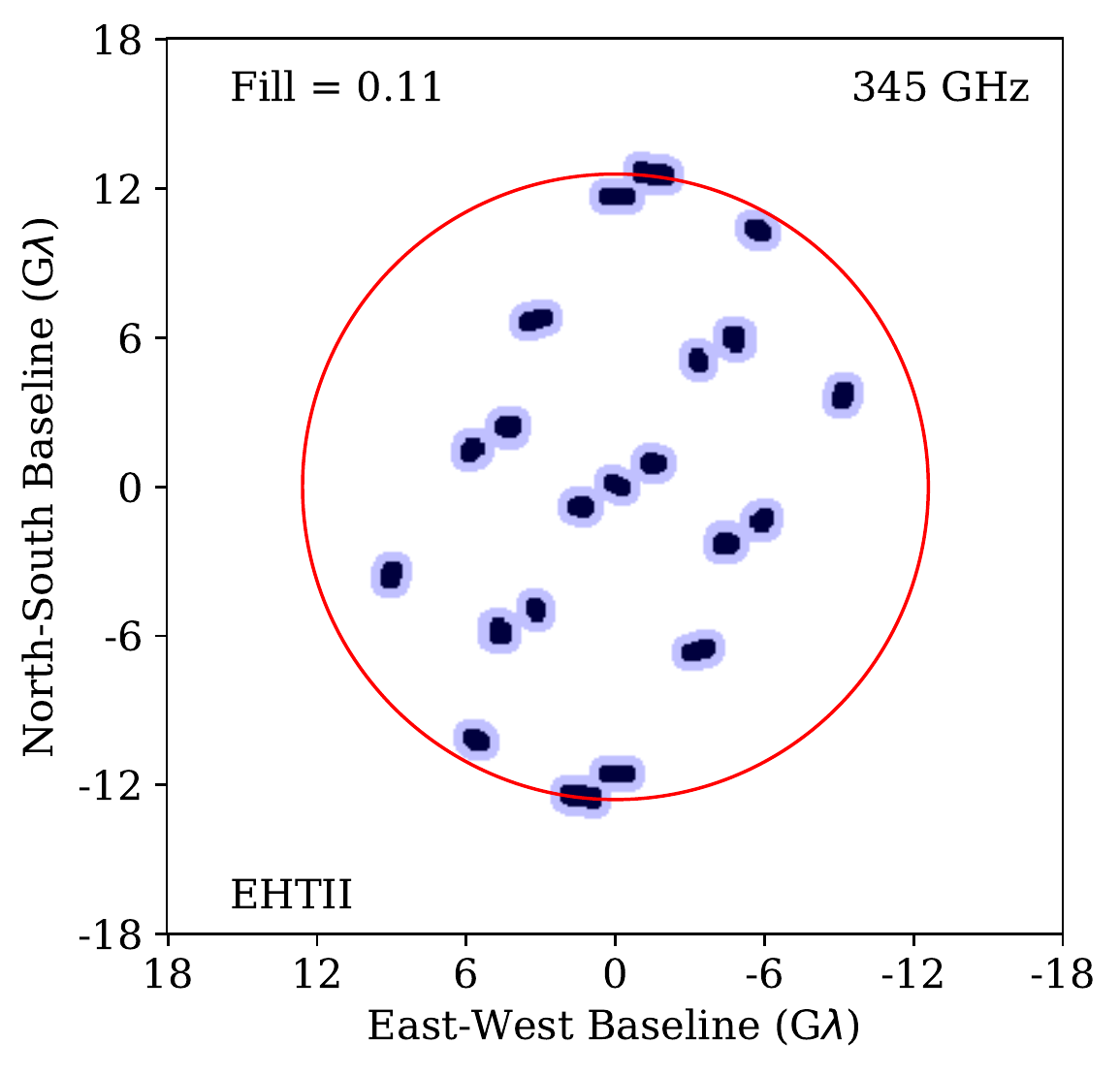}
    \includegraphics[width=.324\linewidth]{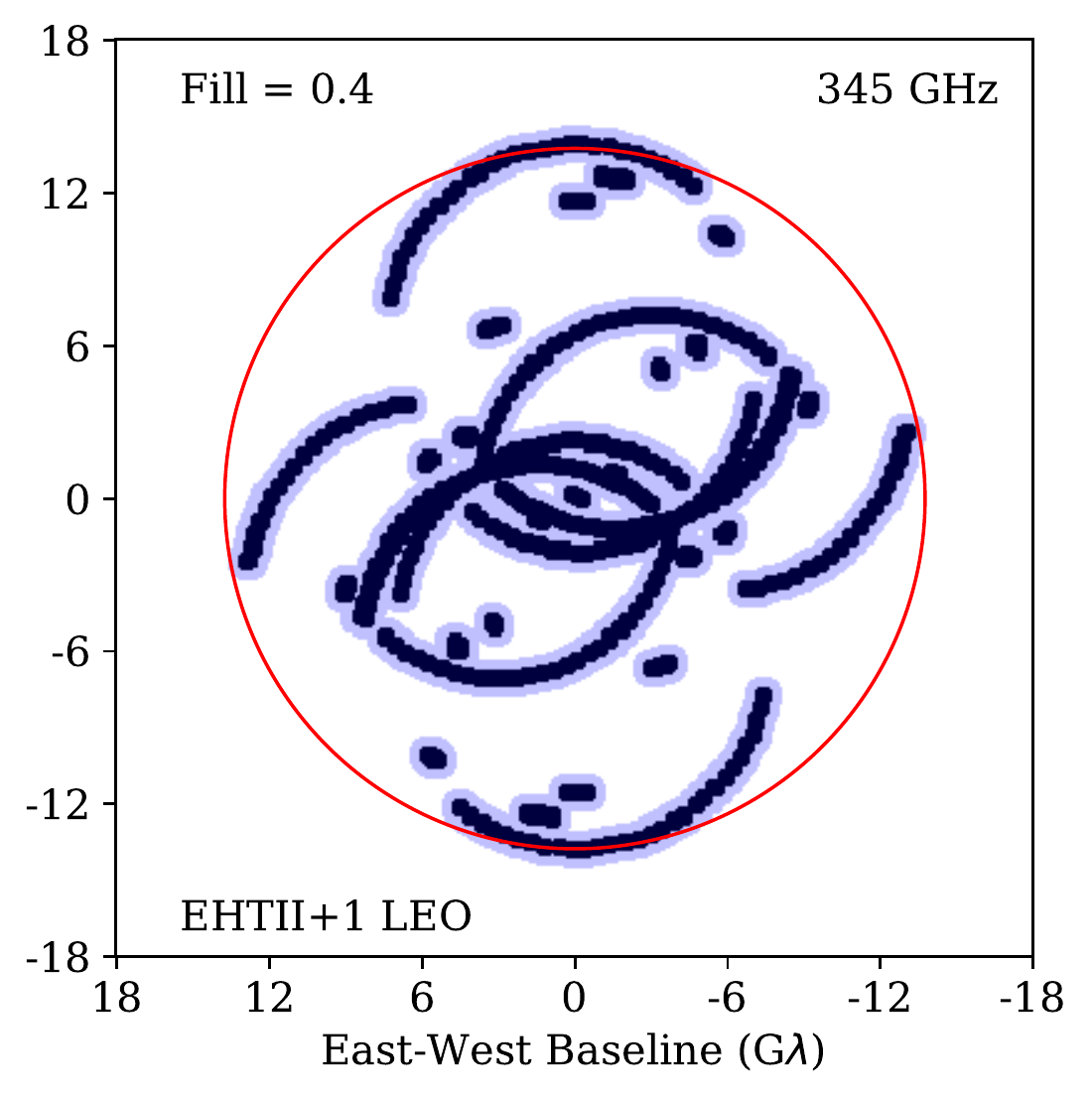}
    \includegraphics[width=.324\linewidth]{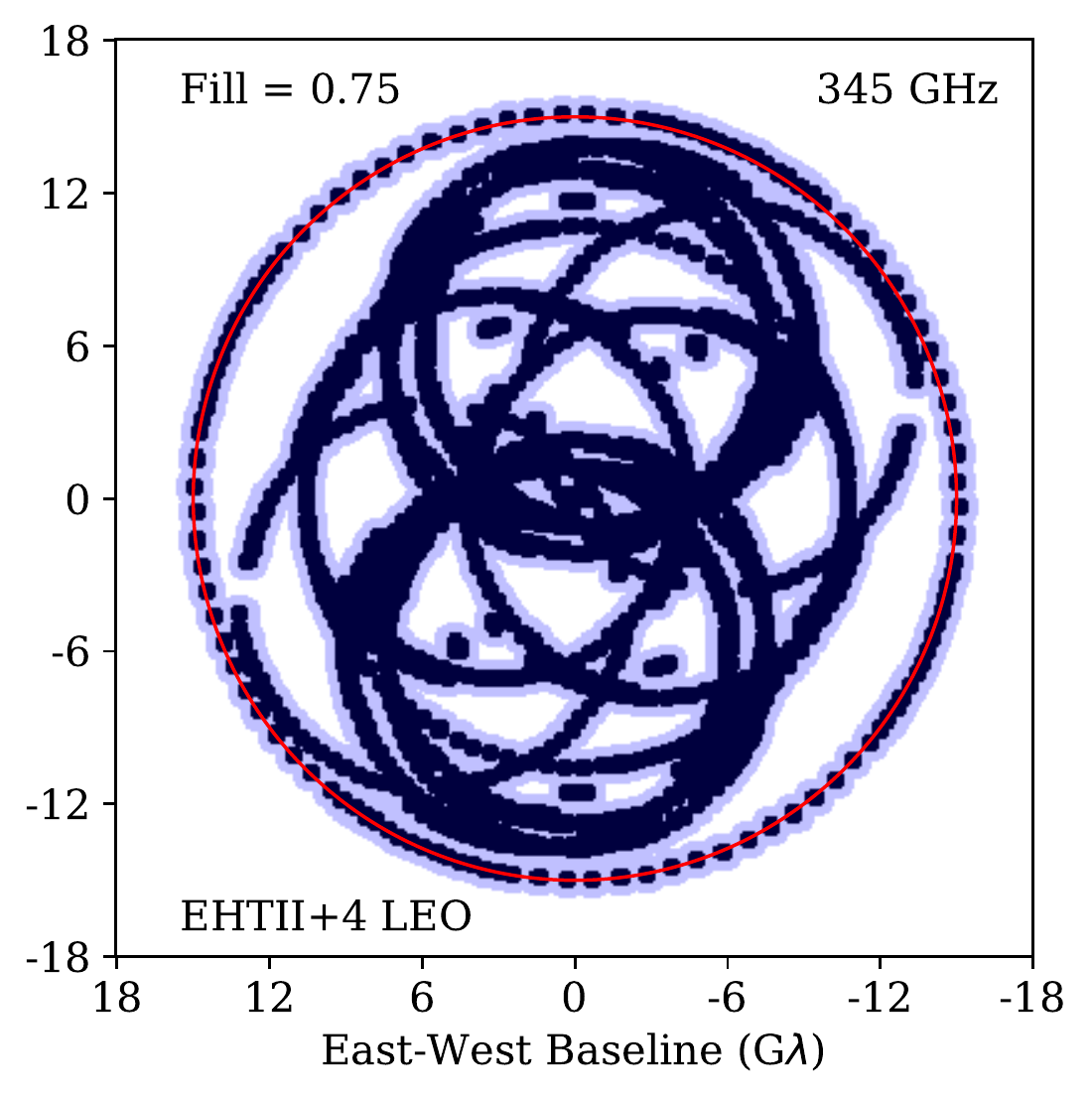}
    \caption{0-0.5 GMST observations of Sgr~A* at 230 (top) and 345\,GHz (bottom) with our $(u,v)$ filling metric applied to the EHTII array, EHTII+1 LEO, and EHTII+4 LEO arrays. The longest baseline circle is shown in red; blue disks show $(u,v)$ coverage convolved with a disk of radius $0.98$ G$\lambda$ (corresponding to a 180 $\mu$as  FOV). Note that the bounding circle, not the convolutional radius in $(u,v)$ space, changes between rows. With four orbiters, the array samples 75\% of the $(u,v)$ plane at a nominal resolution of 12 $\mu$as at 345\,GHz in this 30-minute interval.}
    \label{fig:fill_frac}
\end{figure*}

\begin{figure}[t]
\centering
\includegraphics[width=.49\textwidth]{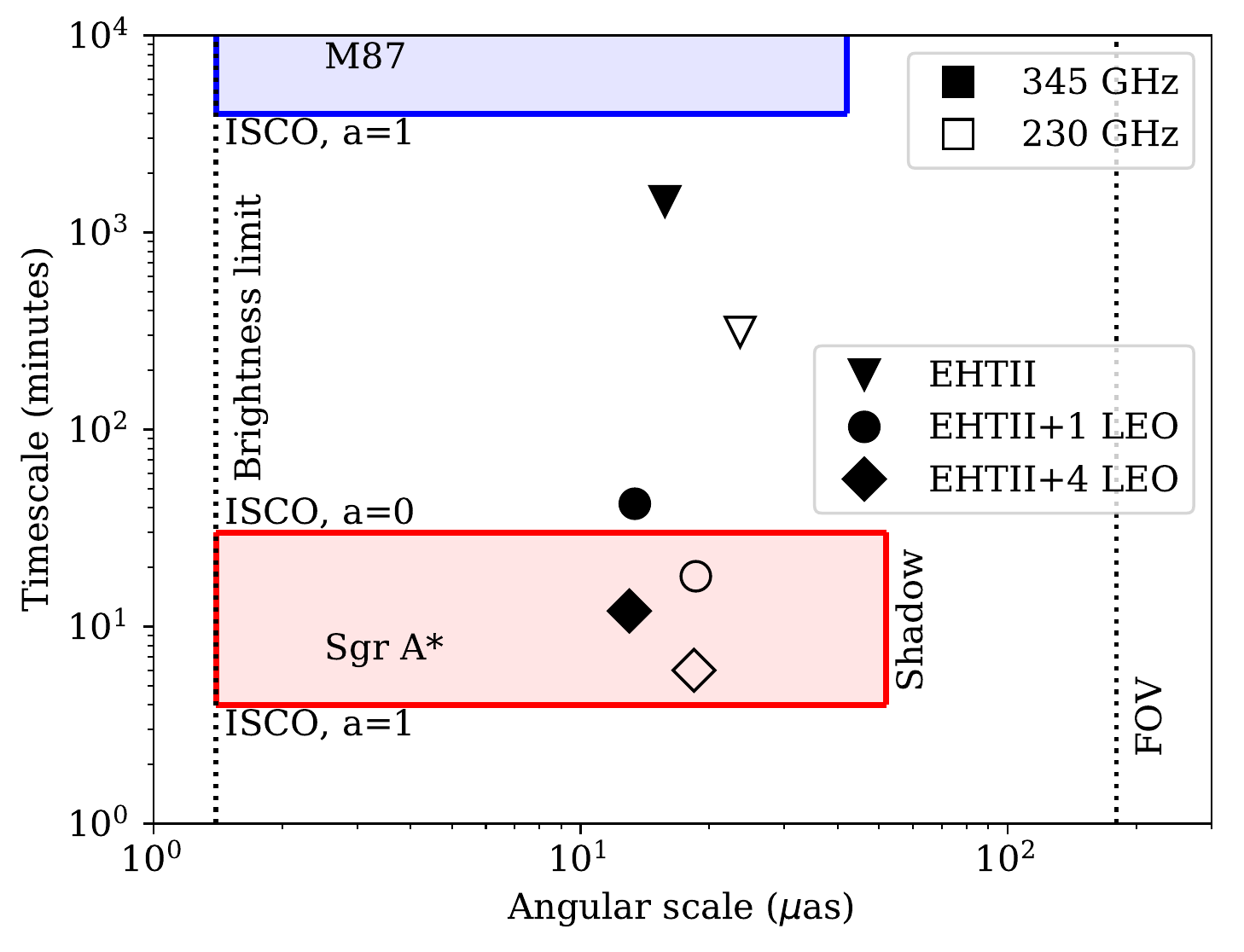}
\caption{Time and angular sensitivities of various arrays, as computed with the FOV-dependent $(u,v)$ filling metric in \autoref{sec:fourier}. The angular and temporal scales of interest for Sgr~A* and M87 are shown in red and blue, respectively. Timescales are computed by finding the shortest observation duration to reach a $(u,v)$ filling fraction of 0.5 for a reconstructed field of view of $180\mu$as. Angular scales are determined by the reciprocal of the longest baseline. $(u,v)$ coverage is dependent on source right ascension and declination; for the filling fractions of each array shown above, coverage is computed for Sgr~A*. Due to the longer baselines at 345\,GHz, high frequency observation has a generally worse $(u,v)$ filling fraction, making imaging more difficult. Space-enabled arrays are required to reach the 50\% filling threshold on timescales sufficient to resolve dynamical features of Sgr~A*. }
\label{fig:scales}
\end{figure}
By rotating, time-delaying, and combining simple circular orbits or orbits of existing space installations, constellations of various orbiting configurations can be created and tested in VLBI simulation environments such as \texttt{eht-imaging} \citep{Chael_2016, Chael_closure}. For this initial examination of LEO imaging capabilities, we use a constellation of orbiters that can always see Sgr~A*, and refer to such orbits as ``face-on.'' Such orbits can be generated for a general source with right ascension $\alpha$ and declination $\delta$ by taking the right ascension of the ascending node $\Omega = \alpha \mp 90^\circ$ and the inclination $i = \delta \pm 90^\circ$, where signs are determined by the handedness of the orbit relative to the source line of sight. We give our orbiters a period of 90 minutes. A diagram of the relative positions and baselines of such an orbit is shown in \autoref{fig:orbit_diagram}. This choice is useful for an initial examination of Earth-space VLBI due to its continuous coverage over time. The total additional $(u,v)$ coverage provided by the orbiter is mostly insensitive to the particular time window used to evaluate coverage; thus, improvements to a dynamical reconstruction can be expected to be approximately constant in time while ground sites can see the target.

However, baselines between space dishes remain constant in length and thus only sweep out concentric circles in the $(u,v)$ plane over repeated orbits. A four-orbiter equispaced paradigm yields only two concentric circles, as opposed to the maximal $n(n-1)/2$ tracks for $n$ dishes. Any expansion of the EHT to contain multiple orbiters would necessarily require a more careful study of orbital configurations. Though a ``face-on'' LEO would decay out of the plane of sight to Sgr~A* after many epochs, the baseline coverage on a timescale of ${\sim}1/2$ a period is representative of the coverage a more general half-Earth-shadowed paradigm would achieve. Moreover, the primary benefits in new coverage arise from space-ground baselines.

Imaging algorithms may benefit from spatial and temporal distributions of $(u,v)$ coverage that are designed to fill specific holes in EHT coverage rather than generically improving total $(u,v)$ sampling over time. Further, the current EHT array is missing short baseline coverage, and an orbit that never crosses the face of the Earth as seen from Sgr~A* would form short baselines primarily with sites that had just come into view; these sites would be looking through the largest possible amount of atmosphere, and thus short-baseline coverage would be less sensitive than for other orbital orientations. Baselines from the ground to a single ``face-on'' orbiter still have thermal noise approximately equal to those between ground sites as shown in \autoref{fig:1ring_cov}, and significantly lower than the long-baseline flux densities of Sgr~A* at 230\,GHz \citep{Lu_2018}, suggesting that even long-baseline observations with orbiters will produce detections.

Incremental changes to the orbiter apogee distance on the order of hundreds of kilometers do not have a great effect on the overall distribution of baselines formed to the orbiter, as such a change is a small fraction of the $(X,Y,Z)$ position vector magnitude of the orbiter, and thus a small change in the $(u,v)$ tracks. 

Though we choose to explore LEOs with dynamical imaging in mind, non-imaging analysis methods would also benefit from the ``face-on'' orbital geometry. Source models can be constrained from the variation of data products alone \citep{Doeleman_2009,Fish_2009,Roelofs_2017}, which would be enhanced by the addition of a LEO. These methods are of particular interest for monitoring of the closure phase, which is the sum of the baseline phases around a triangle of antennas. The baseline from a ``face-on'' LEO to the South Pole Telescope can always see Sgr~A*. Thus, whenever any other ground station can see the source, a closure triangle is formed for the entire duration of that site's observing window, enabling the longest possible monitoring for closure phase variation while still including two ground stations. The SPT and orbiter are relatively insensitive compared to other ground sites; in the case of a three-telescope observation, the third dish would likely need to be ALMA or a similarly sensitive site in order to achieve reliable detections. However, because this orbital configuration may not be feasible for a real expansion of the EHT to space, these potential science targets should be considered as inspiration, not as justification, for a particular orbital paradigm.

Though there is significant freedom in the optimization of orbital elements, we take a ``face-on'' orbiter as an intuitive proxy for the expected performance of a LEO station. We use constellations of phase-shifted ``face-on'' orbiters to test arrays with increasing numbers of space dishes. 

\subsection{Time and Angular Scale Sensitivity}

In order to perform a quantitative and imaging algorithm-independent comparison between potential arrays, we construct a $(u,v)$ filling fraction metric based on the geometric necessity to sample the $(u,v)$ plane with sufficient density to model the source intensity distribution. Equivalently, a single observation in the $(u,v)$ plane constrains the possible visibilities in a region around this point determined by the field of view $\theta_{\textrm{FOV}}$ in accordance with the Nyquist-Shannon theorem as formulated for VLBI \citep{Bracewell_1958}. For a source with finite extent on an otherwise empty sky, the sky intensity distribution is given as a function of the sky position $(\theta_x, \theta_y)$ by the Fourier transform from the interferometric visibility:
\begin{align}
I(\theta_x, \theta_y) =\frac{1}{2 \pi} \int_{-\infty}^{\infty}\int_{-\infty}^{\infty}V(u,v) \exp{[2 \pi i (u \theta_x + v \theta_y)]} du dv.
\end{align}
Realistic baseline coverage is discrete and finite, so the intensity distribution is effectively interpolated between points in the visibility function $V(u,v)$. In the simple case of a filled disk of brightness on the sky, the Fourier transform is a Bessel function, for which the half-width at half-maximum is approximately $0.71/\theta_{\rm FOV}$. If we assume that the imaged source lies entirely within the angular extent $\theta_{\textrm{FOV}}$, then equating the argument of the Bessel function with the computed half-width gives a $(u,v)$ sampling radius as a function of FOV:
\begin{align}
\theta_{\textrm{FOV}} |\vec{u}| = 0.71.
\end{align}
Thus, we find a visibility sampling radius $|\vec{u}| = \frac{0.71}{\theta_{\textrm{FOV}}}$. By convolving the baseline coverage of an observing session with a disk of this radius, we obtain a representation of the visibility function constrained by the observation. The longest baseline in the observation sets an outer radius in the $(u,v)$ plane within which the convolved coverage fills in points. Within this circle we compute a fraction of $(u,v)$ area constrained for imaging purposes at a particular field of view for the nominal resolution of the observing session. A comparison of this metric applied across EHT arrays with increasing orbiters for a source confined to within $180\,\mu{\rm as}$ (corresponding to a $(u,v)$ sampling radius of approximately 0.98 G$\lambda$) is shown in \autoref{fig:fill_frac}. Note that for other simple source models with equivalent angular extent the metric half-width may differ significantly, such as for a pair of point sources separated by $\theta_{\rm FOV}$, which yields a sampling radius $|\vec{u}| = \frac{0.33}{\theta_{\rm FOV}}$.

By creating synthetic observations across starting times and durations throughout a day, we determine the minimal observation duration required to reach a particular filling fraction, effectively finding a minimum timescale for observation that depends only on the target angular extent and position in the sky. The ordered pair of (timescale to filling, nominal resolution) provides a concise pre-imaging comparison tool for array configurations, dependent only upon the assumption of a source FOV. As the fractional filling timescale depends on when observations begin, we compute the optimal start time by comparing combined sampling over time across all start times with a resolution of a tenth of an hour.

\autoref{fig:scales} shows the expected temporal and angular sensitivities for three simulated arrays, both at 230 and 345\,GHz. We define temporal sensitivity as the shortest observation that reaches a filling fraction of 0.5 for a particular source, and we define angular sensitivity as the angular resolution of the half-filling observation. We compare these values with the relevant temporal and angular scales for Sgr~A* and M87. For each source, we assume a conservative FOV of $180\,\mu{\rm as}$, which is a small factor larger than the Gaussian image full-width at half of maximum inferred from previous EHT observations of each source  \citep{Doeleman_2008,Fish_2011,Doeleman_2012,Johnson_2015_science,Akiyama_2015,Lu_2018}. To bracket the representative timescales of each source, we use the ISCO period at zero spin ($P_{a=0} = 12\sqrt{6}\pi t_{\rm G} \approx 92.3 t_{\rm G}$) and at maximal spin ($P_{a=1} = 4\pi t_{\rm G} \approx 12.6 t_{\rm G}$) for the gravitational time $t_{\rm G}= G M / c^3$ and black hole mass $M$. For the maximal representative angular scale, we use the expected Schwarzschild shadow diameter ($2\sqrt{27} r_{\rm G}/D \approx 10.4 r_{\rm G}/D$, where $D$ is the distance from the observer to the black hole and $r_{\rm G} = G M / c^2$). For the minimal representative angular scale, we adopt a physical limit based on a maximum brightness for synchrotron radiation: $T_{\rm B} \lesssim 10^{12}\,{\rm K}$ \citep{Kellermann_1969,Readhead_1994}. Adopting this brightness temperature limit and requiring that image features must be at least $100\,{\rm mJy}$ to be detectable in the array configurations we examine, we estimate that image features must have angular size exceeding $1.5\,\mu{\rm as}$ at 230\,GHz (or $1.0\,\mu{\rm as}$ at 345\,GHz).

Though we have focused our analysis of $(u,v)$ filling on compact emission, the ability to resolve extended structure of compact sources also improves with the addition of LEOs. Though reconstructions of a larger field of view correspond to a smaller region of effective sampling around each $(u,v)$ point and thus a generally smaller fractional coverage, successive orbits of LEO stations provide dense coverage with space-ground baselines due to the relatively slow motion of ground sites. Moreover, we find that the addition of at least one ``face-on'' orbiter is required for the EHT array to temporally resolve a $180 \mu$as FOV Sgr~A* model by reaching a filling fraction of 0.5 on sub-ISCO timescales.

\section{Example Reconstructions}
\label{sec:imaging}
\begin{figure*}
    \centering
    \includegraphics[height=6.6cm]{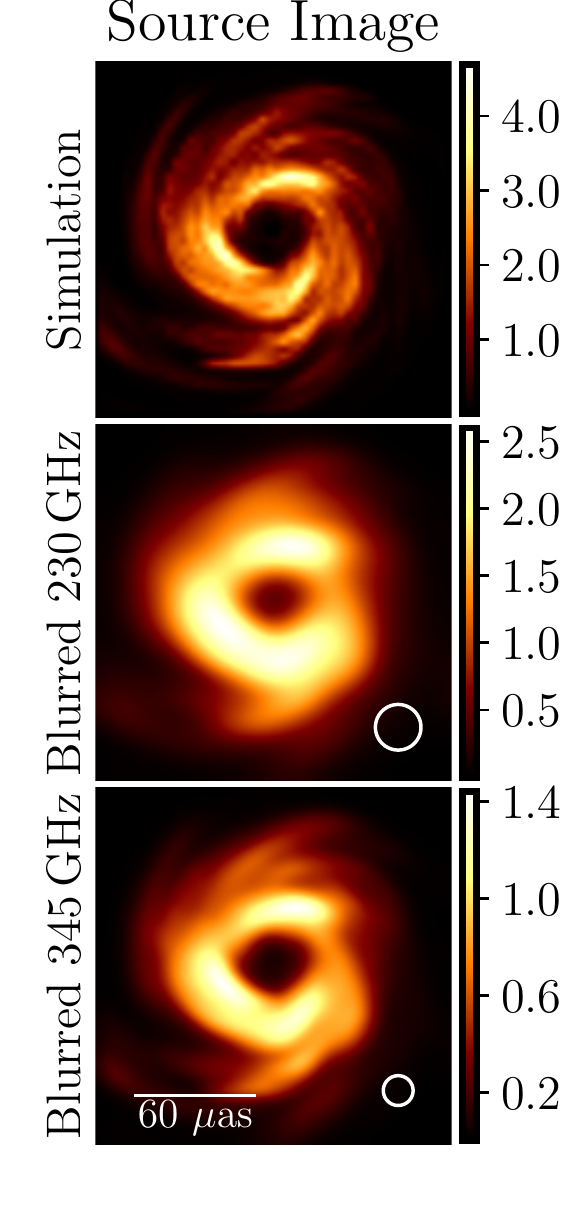}
    \includegraphics[height=6.6cm]{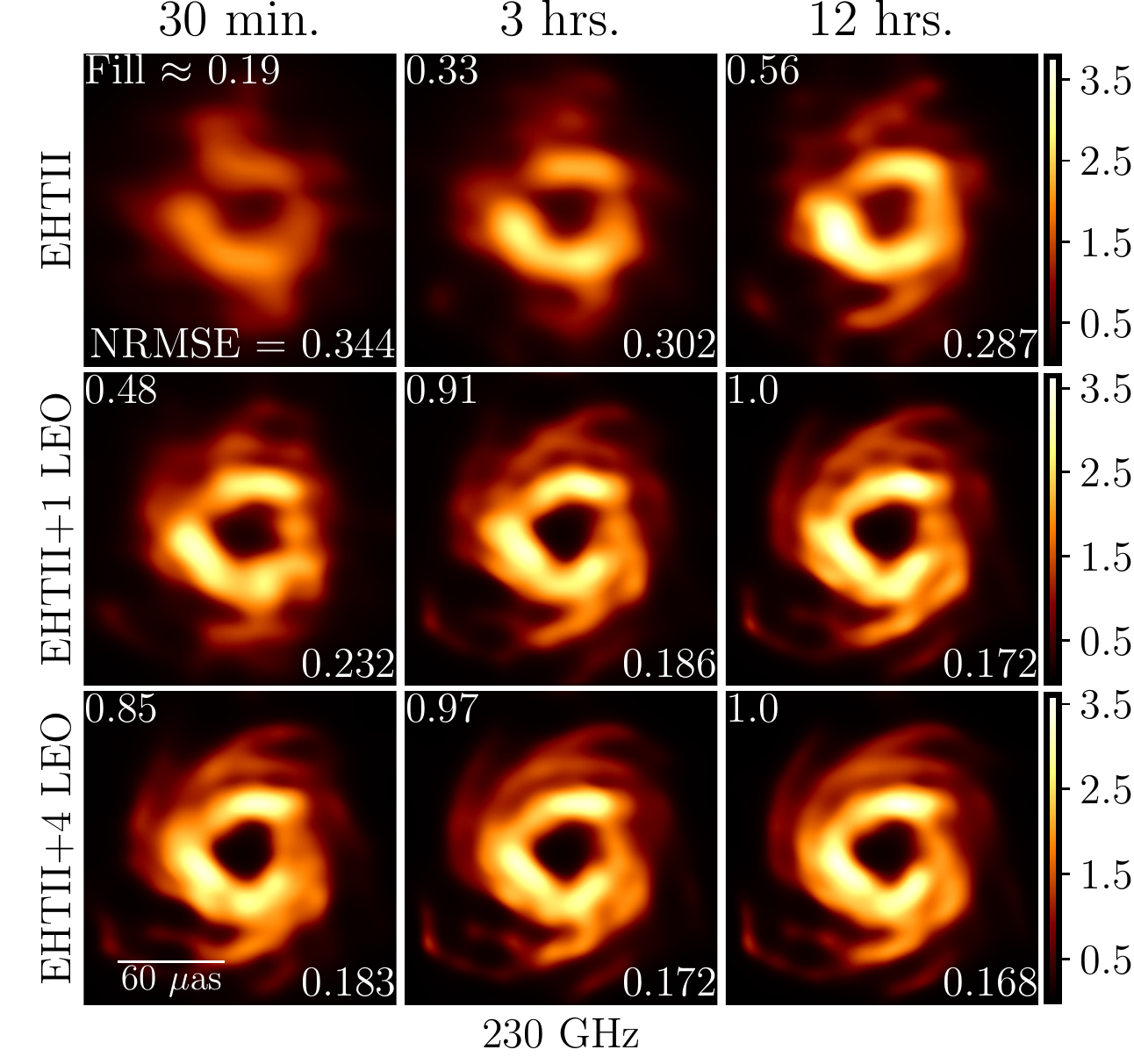}
    \includegraphics[height=6.6cm]{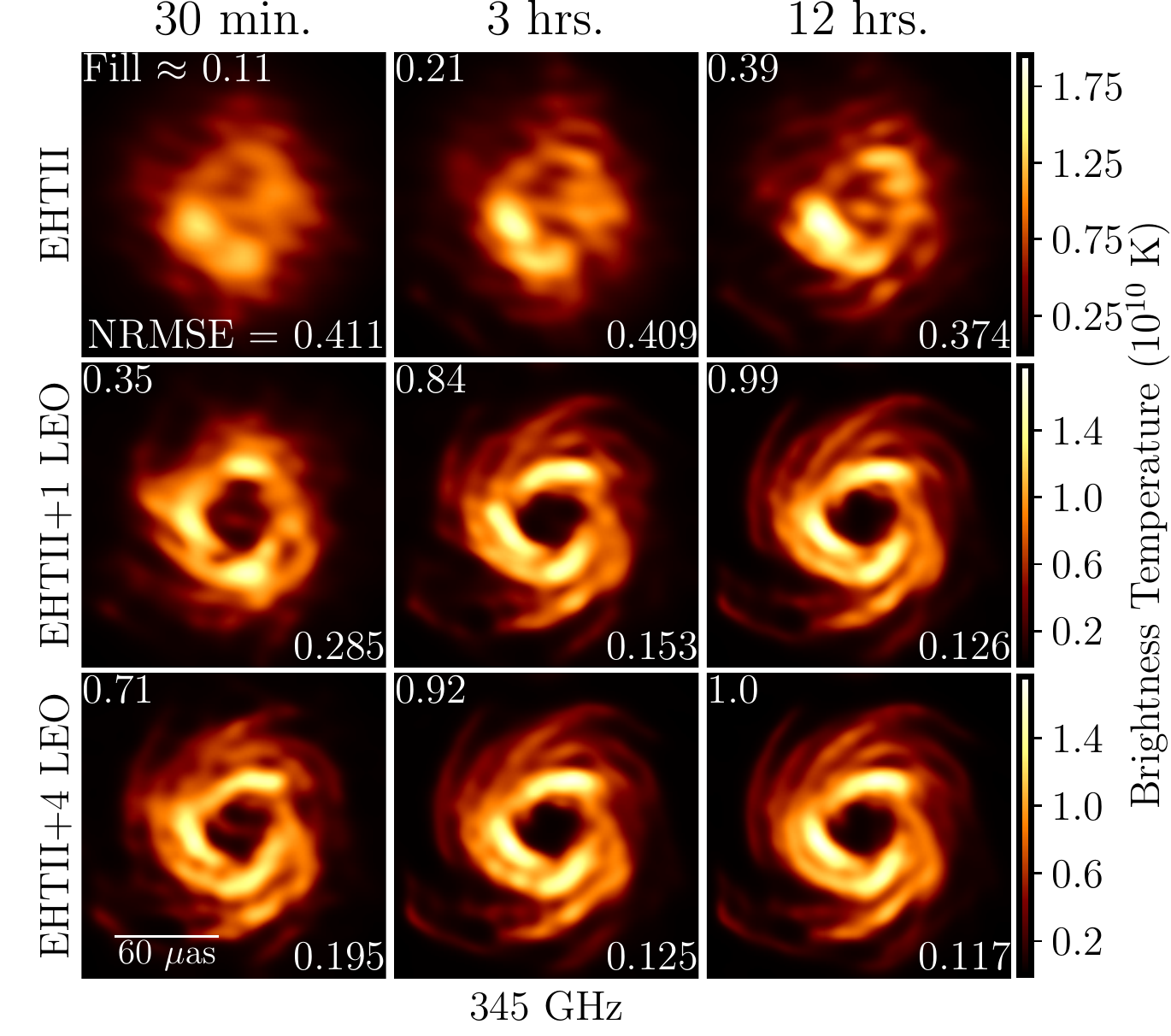}

    \caption{Simulation (left), 230 (middle) and 345\,GHz (right) static imaging of a GRMHD model of Sgr~A* with the EHTII, EHTII+1 LEO, and EHTII+4 LEO arrays for a variety of observation durations centered on 0 GMST. We assume the simulation is totally achromatic between 230 and 345\,GHz such that the flux is constant at each frequency, leading to a difference in the brightness temperature. The brightness temperature at top left is shown for 230 GHz. The source images are shown blurred to half of the nominal resolution of the EHTII array observing Sgr~A* at each frequency (23 and 15\,$\mu$as respectively); imaging typically outperforms the nominal resolution by a small factor. Images are blurred by the ensemble-average interstellar scattering kernel before observation, while imaging scripts attempt to deblur this effect. Normalized root-mean-square error relative to the true image is shown at bottom right in each frame; the $(u,v)$ filling fraction (defined in \autoref{sec:fourier}) is shown at top left. The true image is shown in the top left frame of the right grid in \autoref{fig:dyn_GRMHD}. As expected, higher filling fractions correspond to improved image fidelity. At 230\,GHz, the ground-based array has smaller gaps in coverage and is more successful at reconstructing the black hole shadow with short observations when the $(u,v)$ plane is not well sampled. However, the LEO-enabled arrays have robust sampling at both frequencies, enabling sharper reconstructions at 345\,GHz even on short observational timescales.}

    \label{fig:static}
\end{figure*}

We now apply static and dynamical imaging methods to simulations of Sgr~A* as observed with ground-only and space-enabled EHT arrays. We take a particular sequence of reconstruction steps for each source reconstruction so that differences in the outputs depend only on the observing arrays and not on user fine-tuning. We first validate the $(u,v)$ filling metric by examining static reconstructions from observations of varying durations and filling fractions at 230 and 345\,GHz. We then examine two simulated movies of Sgr~A* that represent various types of time-variability that might be found at the Galactic Center. In all synthetic observations, we include the frequency-dependent ensemble-average blurring effect of interstellar scattering presented in \citet{Johnson_2016} and \cite{Psaltis_2018_scattering} and implemented in \texttt{eht-imaging}. Our imaging scripts deblur observations with the expected diffractive kernel at the frequency of observation \citep{FIsh_2014}.

\subsection{Imaging Methods}

We use the regularized maximum-likelihood imaging methods implemented in \texttt{eht-imaging} to reconstruct static images from synthetic VLBI data \citep{Chael_2016,Chael_closure}. For the purposes of our comparison, we use an identical script across frequencies and arrays, primarily using maximum-entropy regularization with a Gaussian prior. Though static imaging does not attempt to find varying structure, the success of imaging of short observation durations provides a simple proxy for the time resolution of an array at particular angular scales. Dynamical methods generally outperform short-duration static imaging in reconstructions of evolving sources due to the smooth sharing of data over time. However, we include static imaging due to its algorithmic simplicity and relative insensitivity to fine-tuning of the imaging script.

``Dynamical imaging'' describes a method of creating a movie from time-separated VLBI data through refinement of successive snapshots. Differences between these snapshots are constrained by the source dynamical timescale and continuity considerations to ensure smooth flow that captures the intrinsic source variability. For the purposes of our exploration of dynamical reconstructions, we rely on two recent algorithms: StarWarps \citep{Bouman_2018} and Dynamical Imaging \citep{Johnson_2017} (hereafter referred to as J17 to avoid confusion with the general term ``dynamical imaging''), both of which are implemented in the \texttt{eht-imaging} software library. Each package takes an ordered list of initialization images, typically centered circular Gaussian flux distributions, and fits a reconstructed image list to observed data using an image prior (also typically a circular Gaussian). Both methods connect inferences across time, allowing observations on timescales longer than the source dynamical timescale to be simultaneously used for imaging.

StarWarps models the VLBI measurements using a Gaussian Markov Model. 
Due to the simplicity of the Gaussian prior and likelihood models used, a closed-form solution to the likelihood maximization exists and produces reasonable results even in the case of significant missing data. 
A belief propagation optimization method, similar to Kalman filtering and smoothing, is used to recover the movie. This method can also be joined with an Expectation-Maximization approach to simultaneously recover an underlying flow field that is assumed to be constant in time. The flow field represents a static model for frame-to-frame evolution that maps each pixel to a directional change in flux per frame, effectively visualizing average motion in the image.

J17 utilizes regularization over a series of images, enforcing heuristics (e.g., image smoothness) that are expected to apply to the accretion flows expected near black holes. The implementation of J17 in \texttt{eht-imaging} includes most of the regularization tools built for forming static images from sparse Fourier data, with the added means of sharing information across time with regularization for smooth variation between adjacent frames, and adherence to an overall flow field. For the ``hot spot'' reconstruction shown in this paper, no flow field regularization is used in either the StarWarps or J17 results, instead favoring simple smoothness regularization.

\subsection{Imaging with Complex Visibilities}

For simplicity, we produce reconstructions using observed complex visibilities despite the fact that the current operating mode of the EHT does not provide absolute phase calibration. We do not include a systematic error budget on complex visibilities. This choice is optimistic but removes complexity from the problem; antenna-based errors depend on the calibration methods used in data reduction and have effects on resulting images that depend highly on the imaging method used.  Complex visibilities are related linearly to the source intensity distribution by a Fourier transform and contain absolute phase information; thus, complex visibilities provide stronger analytical constraints in the imaging process than closure phase, which is non-linear in source intensity. Using this data product does not represent more measurements in the $(u,v)$ plane; instead, it represents knowledge of a phase reference for each dish in the array at all times. This information in turn decreases the total degrees of freedom while imaging, constraining the reconstruction even if absolute astrometry is not required.

Obtaining absolute phase information for the future EHT is conceivable but nontrivial, requiring calibrator observation quasi-simultaneous with the target, phasing to other dishes in the network, atmospheric characterization, or some other method. Further, absolute phase calibration has been demonstrated at mm wavelengths \citep{Rioja_Dodson_2011}. If analysis of the requirements for absolute phase information show that is not likely to be achievable, then further study of the phase-uncalibrated imaging capabilities of space dishes (with, e.g., visibility amplitudes and closure quantities) will be necessary, though the comprehensive coverage provided by space dishes will likely compensate for the loss of phase information in closure imaging techniques such as in \citet{Chael_closure}.

\subsection{Imaging Sgr~A*}

\begin{figure}[t]
    \centering
    \includegraphics[width=.49\textwidth]{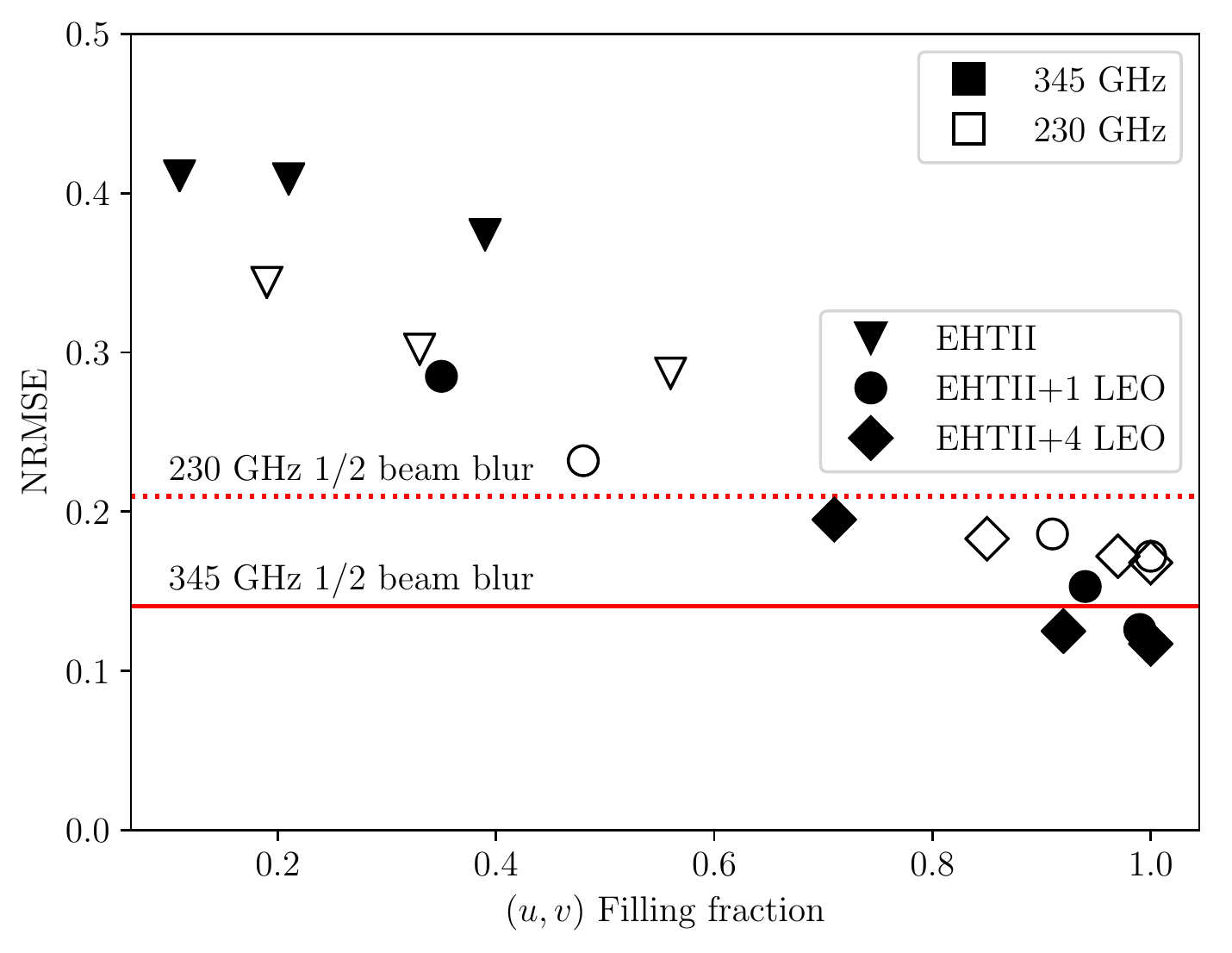}
    \caption{Comparison of normalized root-mean-square error of image reconstructions with the associatd $(u,v)$ filling fraction of the synthetic observation. Data is equivalent to that shown in \autoref{fig:static}. Horizontal lines show the NRMSE of the simulated image relative to itself after convolution with a circular Gaussian of half the nominal resolution of the observing array at each frequency. Image accuracy varies approximately linearly with filling fraction as each array saturates the $(u,v)$ plane with longer observations, until an apparent plateau at the NRMSE achieved by the half-beam blur.}
    \label{fig:nrmse_fill}
\end{figure}

\begin{figure*}
    \centering
    \includegraphics[width=.49\textwidth]{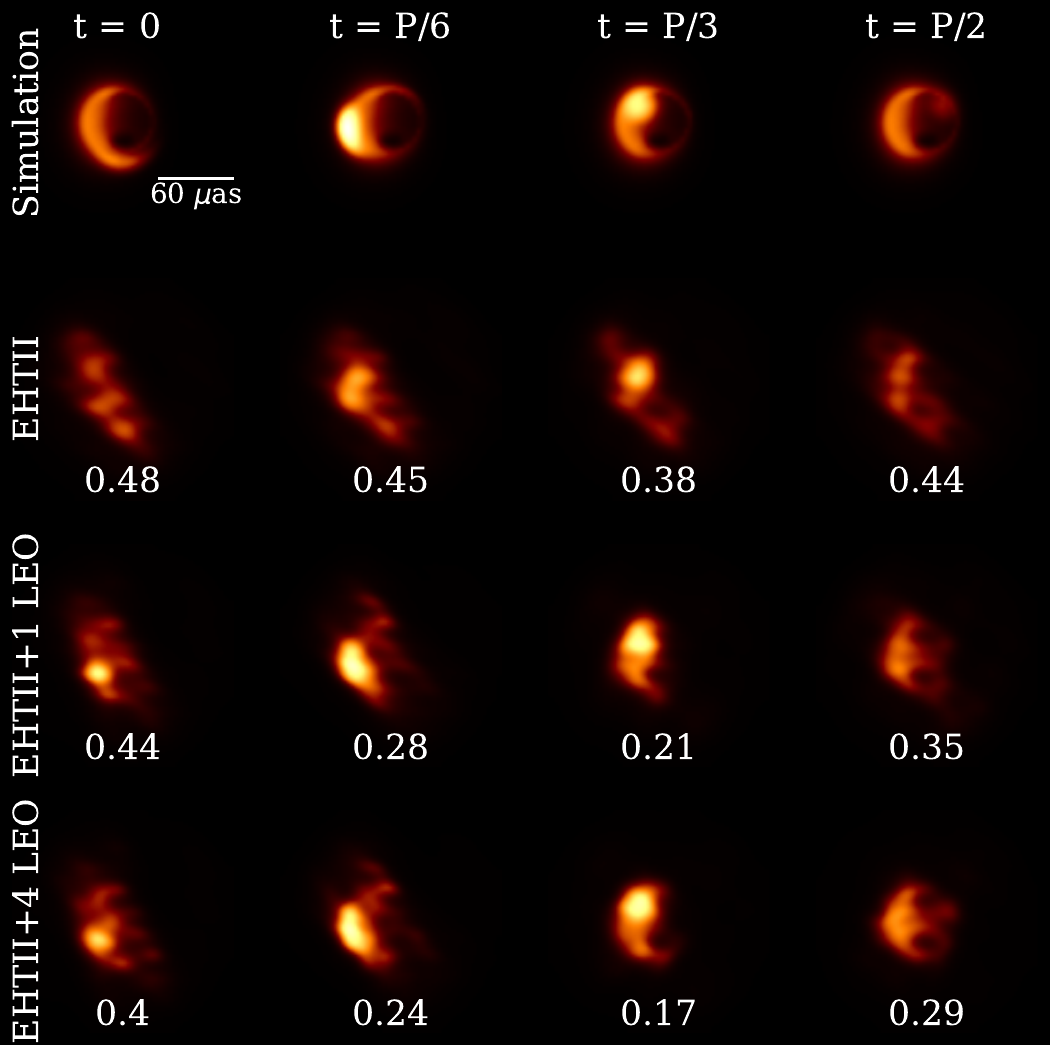}
    \includegraphics[width=.49\textwidth]{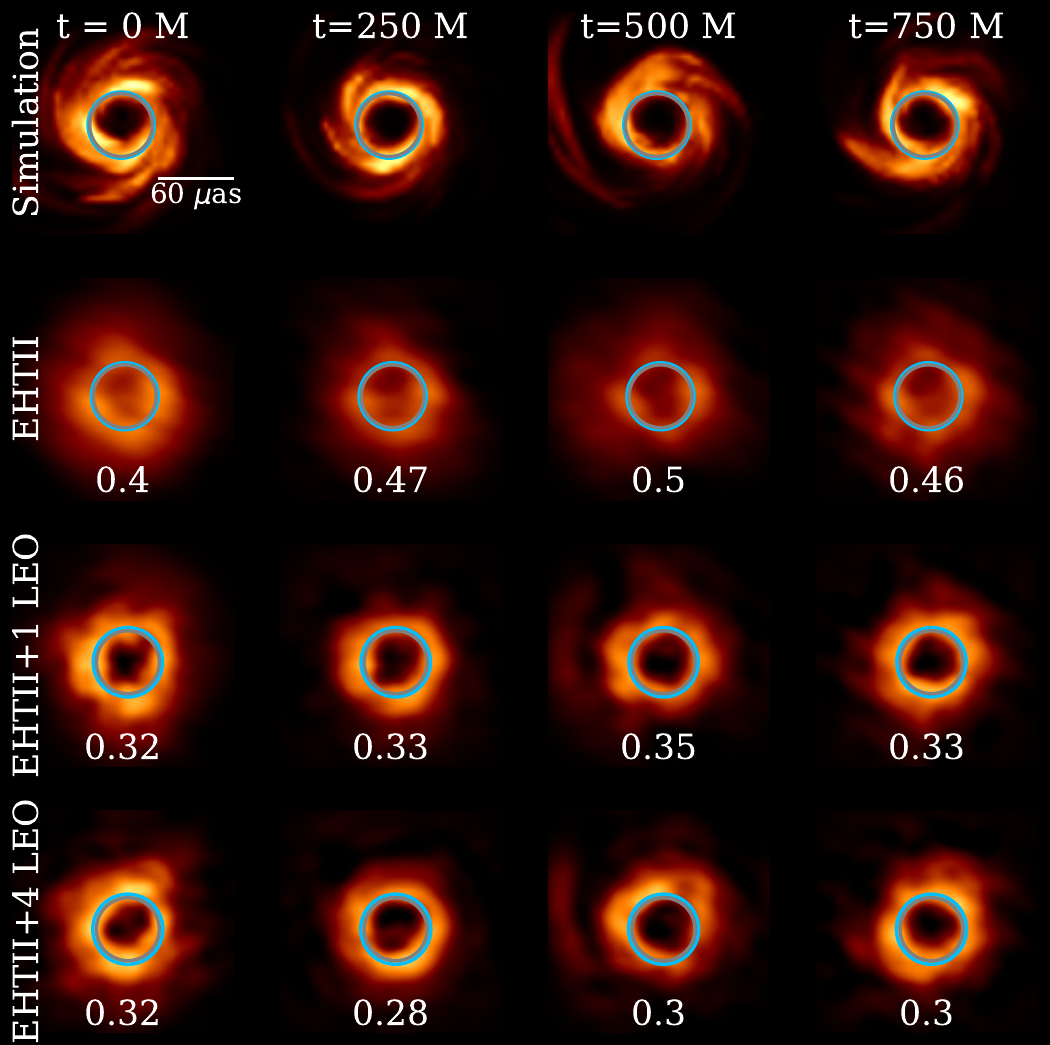}
    \includegraphics[width=.49\textwidth]{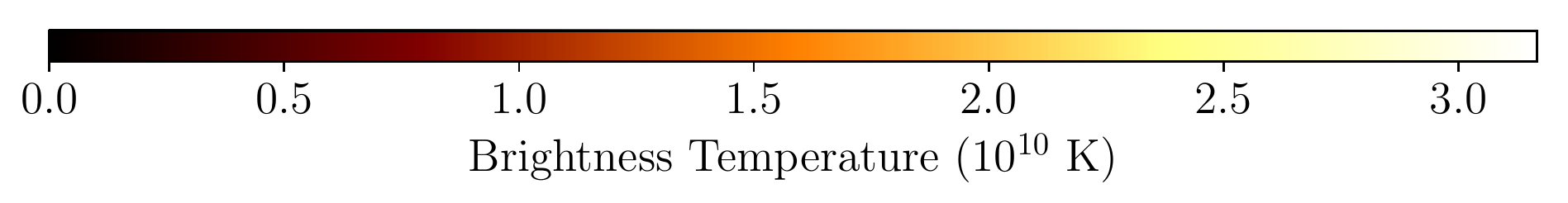}
    \includegraphics[width=.49\textwidth]{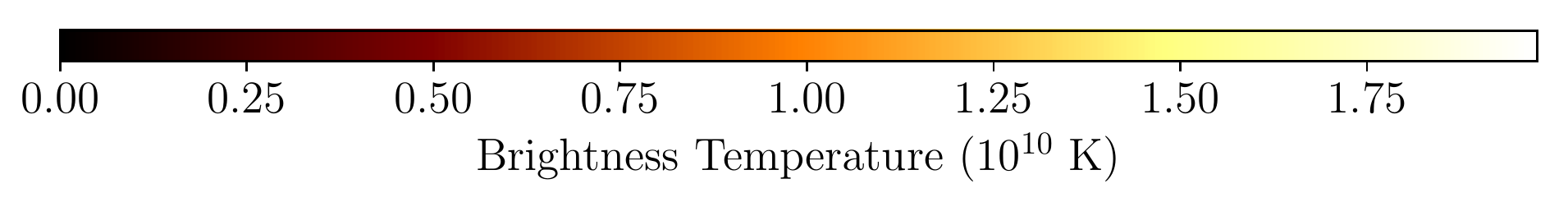}
    
    \caption{345\,GHz dynamical imaging of two simulations. At left, a ``hot spot'' simulation with a 30 minute orbital period is observed over 2 hours starting at 0 GMST with an integration time of 30 seconds and $50\%$ duty cycle, using a 16\,GHz bandwidth. A movie is reconstructed with J17, for which reconstructed frames are shown every sixth of a period. Images are blurred by the ensemble-average interstellar scattering kernel before observation, while imaging scripts attempt to deblur this effect. The NRMSE between the reconstructed frames and the input frames is shown below each image. At right, we show a Starwarps reconstruction of a zero spin $4{\times}10^6\textrm{M}_\odot$ GRMHD simulation at Sgr~A*. Observational parameters are effectively identical to those used for the ``hot spot'', but occur over 6 hours, beginning at 21 GMST. A simple ring fit is performed on the average image of each GRMHD reconstruction and plotted on each image in blue, while the physical photon ring diameter of 51 $\mu$as is shown in gray. Note that the ring fitting algorithm used here overestimates the ring diameter when applied to the truth movie. Adding orbiters allows the finer extended features to be resolved, particularly visible in the spiral arms at the edges of the reconstruction, and leads to a more precise (but not accurate) ring fit. The average image ring fit diameters of the truth movie, the ground-only reconstruction, 1-orbiter reconstruction, and 4-orbiter reconstruction are $53.7\pm1.3\mu$as, $54.4\pm3.1\mu$as, $56.3\pm2.4\mu$as, and $56.9\pm1.5\mu$as, respectively.}

    \label{fig:dyn_GRMHD}
\end{figure*}

Reconstructing static images of Sgr~A* at the event horizon scale has been the focus of much of EHT imaging algorithm development. The current EHT array is expected to be capable of reconstructing static images of Sgr~A* \citep{Doeleman_2009_decadal}. To elucidate the effect of increasing $(u,v)$ filling, and to examine the difference among arrays in the capability of imaging fine structures at a large FOV, we image the first frame in a GRMHD simulation of Sgr~A* that features prominent spiral structure out to a field of view of $180\,\mu$as \citep{Chael_2018}. However, Sgr~A* is expected to evolve rapidly in time, with an innermost stable circular orbital timescale of less than half an hour. We thus also explore how well space-enabled arrays and existing dynamical imaging techniques work for two simulations of variability of Sgr~A*. We use the normalized root-mean-square error (NRMSE) to compare reconstructions to true images, computing pixel-wise RMS differences in aligned images and normalizing to the total flux of the true image.

First, we examine the first frame of a simulation \citep{Chael_2018} of a $4{\times}10^6 \textrm{M}_\odot$ black hole with spin 0 observed at $10^\circ$ inclination. We simulate observations with a $50\%$ duty cycle (integrating half of the total observation duration), 30\,s integration time, and 16\,GHz bandwidth at 230 and 345\,GHz to examine the effects of resolution (tuned by frequency) and filling fraction (turned by observation duration). The ground-based array used in these simulations is the full ``EHTII'' array, including KP and NOEMA. As is shown in \autoref{fig:static}, the LEO-enabled EHTII succeeds in reconstructing fine spiral structures with a ${\sim}5\,\mu$as scale out to a $180\,\mu$as FOV with only 30 minutes of observation. We also see the expected pattern of improvement with increasing $(u,v)$ filling, as well as the difficulty of the ground-based array in the transition to 345\,GHz due to larger unsampled regions in the $(u,v)$ plane when imaging a large FOV.  Moreover, long observations with a space-enabled array saturate the sampling of the $(u,v)$ plane for a static image, so 345\,GHz reconstructions overtake 230\,GHz reconstructions in accuracy.

This transition is visually apparent in \autoref{fig:nrmse_fill}, in which the static reconstruction NRMSEs are plotted against the $(u,v)$ filling fraction. NRMSE decreases with increased filling fraction until the $(u,v)$ plane is well-sampled. Large differences in NRMSE at the same filling fraction occur primarily at low filling fractions between ground and space-enabled arrays; in these cases, the structure of the baseline coverage is likely dominant, indicating that our metric does not fully capture the differing benefits of additional coverage in different unsampled regions of $(u,v)$ space. However, the broad trends behave as expected, including a plateau of NRMSE near the half-beam-convolved level expected for a well-sampled $(u,v)$ plane at each frequency.

As treated in \citet{Broderick_2006}, a ``hot spot'' in orbit around a black hole provides a useful model for intense time variation at Sgr~A*, and may have been observed via polarization time-variability by \citet{Gravity_2018_orbit}. We simulate observations of a hot spot with a 30 minute orbital period from 0-2 GMST and reconstruct movies of the motion with data from the same three arrays as are used for the static reconstructions. The observations have an integration time of 30 seconds observed every 60 seconds. J17 reconstructions of a hot spot using observations at 345\,GHz are shown at left in \autoref{fig:dyn_GRMHD}; the addition of space-VLBI stations is required to resolve the ``hot spot'' feature moving across the dimmer constant image. NRMSE sharply improves with the addition of one orbiter, while four orbiters show less drastic additional improvement. The fall-off in improvement with additional orbiters can be partially attributed to the orbital equispacing of the four-orbiter case; separating each orbiter by a different distance would improve baseline coverage, though an optimization of separation is beyond the scope of this paper.

General relativistic radiative magnetohydrodynamic (GRRMHD) simulations provide a more realistic picture of what might lie at Sgr~A*. We apply a StarWarps imaging pipeline to the full simulation corresponding to the single frame shown in \autoref{fig:static}. Reconstructions using observations with a ${\sim}60\%$ duty cycle but otherwise identical parameters to those used for the ``hot spot'' are shown in \autoref{fig:dyn_GRMHD}. The change in duty cycle results from a minimum time separation in frames of the simulation of $10 t_{\rm G} = 197.1$\,s during which we perform four 30\,s integrations. Due to the degeneracy of the effect of accretion disk orientations \citep{Broderick_2011_symmetry,Johnson_2015_science} upon the black hole shadow geometry, temporally resolving this source model to determine the flow direction (clockwise or counterclockwise) is of particular interest to the EHT.

Black holes are expected to exhibit a bright ring of emission corresponding to photon trajectories that orbit the black hole before escaping \citep{Cunningham_1976,Laor_1990,Viergutz_1993,Bao_1994,Cadez_1998,Agol_2000,Beckwith_2005}. This ring is largely unaffected by accretion dynamics and instead relies primarily on black hole mass and spin \citep{Bardeen_1973, Johannsen_2010}. We can measure the shadow size from reconstructed images by performing a simple ring fit to reconstructed images by finding the ring center and profiles via brute force search, minimizing the standard deviation of distances from the ring center to the next brightness peak along many angular slices \citep{Chael_rings}. This algorithm is applied to reconstructions of the GRMHD simulation in \autoref{fig:dyn_GRMHD}.

Without space dishes, the ground-based array fails to find the detailed features of either evolving model. In particular, the ground reconstructions cannot reliably extract the shadow in the GRMHD simulation. By contrast, the space-enabled arrays are capable of reconstructing both the motion of the hot spot and the larger extended structures in the GRMHD simulation. However, the accuracy of the resulting ring fits are not linearly related to the fidelity of the image reconstructions. In this particular simulation, this is likely because the space-enabled reconstructions resolve bright features beyond the ring which push the fit further out from the physical radius. Further, the precision of the ring fits do not fall below the $\pm4\%$ sensitivity nominally required to measure spin from a shadow measurement \citep{Bardeen_1973,Johannsen_2010}. This precision requirement ($\ll4\mu$as for a ${\sim}50\mu$as lensed photon ring) is unsurprisingly difficult to surpass even with a space array; the required angular resolution is far below the diffraction-limited resolution of the arrays we consider. Moreover, the space-based array reconstructions will allow individual tracking of evolving features around a well-resolved shadow; thus, algorithms focused on parameter estimation by tracking matter orbits in the image domain will enable better measurements of spin.

\section{Discussion}
\label{sec:discussion}

We have developed tools to simulate  observations and imaging with VLBI arrays that include both ground-based and orbiting dishes within the open-source EHT codebase \texttt{eht-imaging}. We have outlined generic constraints on space-VLBI that will inform any future consideration of a space-enabled array. We have implemented a $(u,v)$ coverage metric that characterizes the temporal and angular sensitivity for a VLBI array, and we have used these tools to analyze the addition of 1 to 4 LEO dishes with 4 meter diameter to the EHT array. We have found that the improved $(u,v)$ coverage of a single orbiting dish enables dynamical imaging on short timescales, resolving changes in structure over less than 30 minutes.

\begin{figure}[t]
\centering
\includegraphics[width=.49\textwidth]{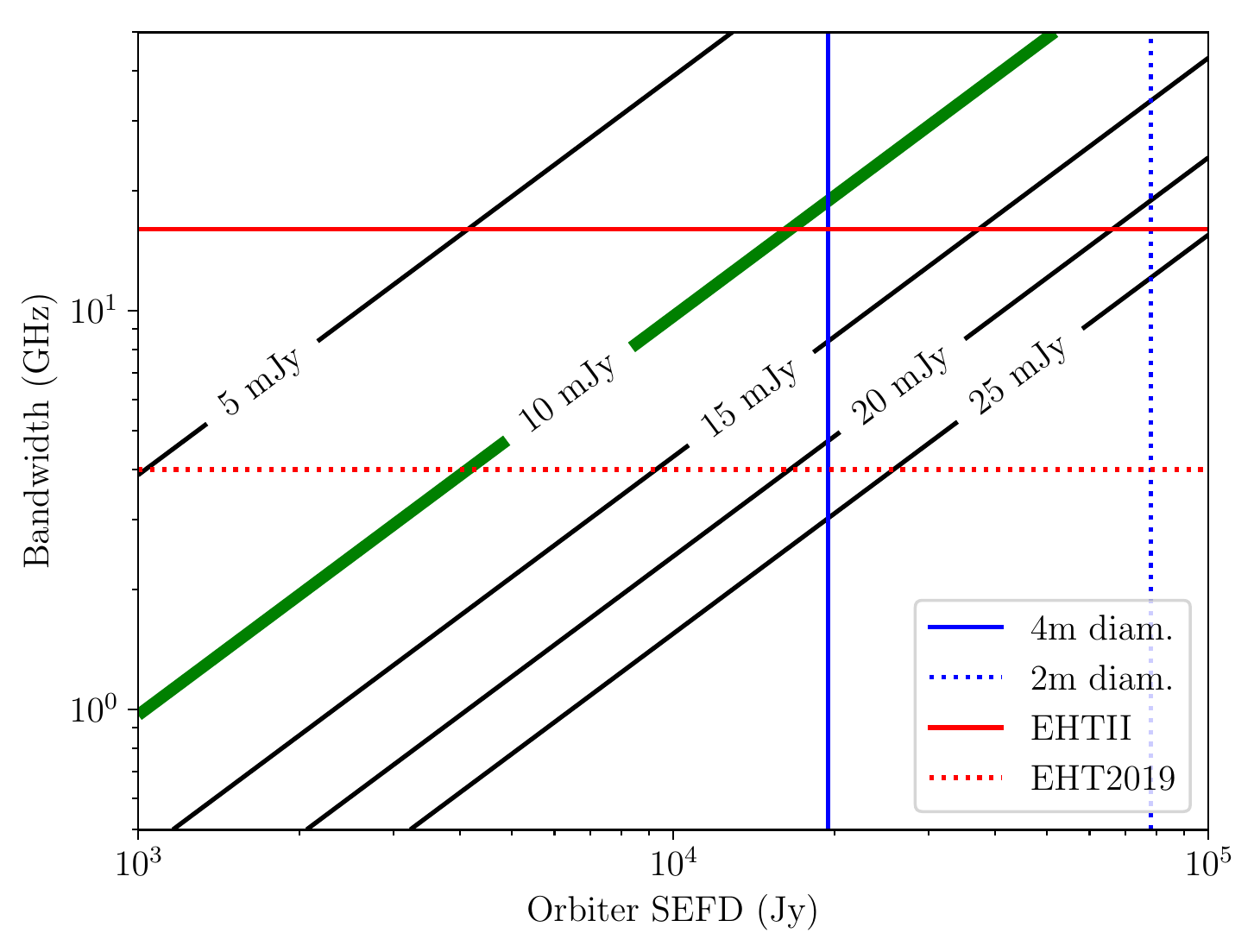}
\caption{Thermal noise contours on baselines from ALMA to a simulated orbiter with an integration time of 1 second. The high sensitivity of space-ALMA baselines on very short timescales indicates that other space baselines can reliably be calibrated by phase steering to ALMA. The region with higher bandwidth and lower SEFD thus constitutes a parameter space for the orbiter with reliable strong detections to ALMA in 1 second of integration, well below the coherence time threshold from orbital constraints found in \autoref{sec:background}. Notably, dish sensitivity and bandwidth can be exchanged to move along a contour, indicating a flexibility in the underlying hardware requirements. In particular, the wide bandwidth of the EHT backend, if transferable to space, makes finding detections to smaller dishes feasible.}
\label{fig:sigma_contour}
\end{figure}

Our paper has focused on assessing the imaging capabilities of potential space-ground VLBI arrays. We have not formulated specific hardware needs for a potential LEO VLBI station operating with the EHT. However, we have identified ${\sim}10$\,mJy as a target thermal noise based on the long-baseline flux observed at Sgr~A*. This value is not a strict boundary for a successful orbiter; further, our estimates of achievable dish SEFD may prove optimistic. Moreover, a compromise in SEFD due to dish size, aperture efficiency, or receiver temperature could be offset by an increase in bandwidth to preserve overall sensitivity (see \autoref{fig:sigma_contour}). Continued EHT studies of Sgr~A* will clarify hardware priorities for a space expansion. 

Enabling time-domain analysis of Sgr~A* is particularly important due to fundamental difficulties in extracting black hole parameters from static reconstructions that are not removed even by perfect reconstructions of the source image. Although the black hole spin is difficult to extract from the shape of the black hole shadow \citep{Johannsen_2010}, it may be tractable to extract spin from a measurement of periodicity near the event horizon, as is done for analysis of quasi-periodic oscillations of X-ray binaries \citep{Ingram_2011, McClintock_2011}. Though intrinsic variation may be mitigated under some conditions, dynamical imaging may be necessary for basic accuracy in reconstructions of the black hole shadow under conditions of intense time variability \citep{Lu_2016,Johnson_2017,Bouman_2018}. Developing robust time-domain analysis tools for sparse VLBI data will thus be required for a serious appraisal of a time-domain-science driven space-VLBI station. The reconstructions shown in this paper do not measure the shadow precisely enough to distinguish black hole spin, indicating the necessity of direct measurements of evolution. Methods that extract periodicty from or fit models directly to variation in the data have been demonstrated on simple time-varying models, and should be generalized to extract spin under broader variational conditions \citep{Doeleman_2009,Fish_2009,Roelofs_2017}. More model-independent methods (e.g., imaging) will be required for analyzing complex or non-periodic evolution.

Studies of other sources will also benefit from the improved coverage of a LEO-enabled EHT regardless of specific orbital geometry. Though other sources do not receive uninterrupted viewing from dishes in the orbital plane facing Sgr~A*, such dishes still form space-ground baselines over at least half of all observing time, providing a rapid increase in $(u,v)$ coverage. Other black hole candidates such as M87, 3C279, and Centaurus~A are not expected to exhibit time variability as rapid as that of Sgr~A*, but swiftly-formed dense coverage still leads to high-fidelity imaging. LEO dishes also benefit reconstructions of extended structure due to the high density of points sampled in the $(u,v)$ plane; reconstructions of extended dynamics would elucidate possible inflow and outflow behavior at Sgr~A* or jet-launching structure at M87. However, extended structure (such as the jet at M87) is likely much dimmer than shadow-scale structure, and so space baselines may not be sufficiently sensitive to achieve long-baseline detections in the small-dish paradigm.

Other work has suggested a space-VLBI array involving two dishes in offset orbits with space-space baselines designed to sweep through broad and regularly-spaced $(u,v)$ coverage \citep{RoelofsFalckeBrinkerink2019a}. This alternate space-VLBI approach could produce high-fidelity static images, but not the rapidly-evolving dynamical movies targeted in the present work. Other expansions to the EHT have been explored, including Medium Earth Orbit (MEO) and Geosynchronous Earth Orbit (GEO) dishes for increased angular resolution \citep{Fish_space}; such expansions would likely be fully complementary with an expansion to LEO, but would not provide comparable short-timescale temporal sensitivity. Balloon-based VLBI may address temporal sensitivity in a similar manner to the LEO orbits we consider; technical feasibility studies that may be transferable to LEO VLBI design are already underway \citep{Doi_2019}. Finally, Spektr-M, or Millimetron, may provide sensitivity at the high frequencies of the EHT in the temporal regimes of relevance to Sgr A* if it is placed in LEO \citep{Millimetron_science}.

While the face-on orbits considered in this paper provide continuous coverage of Sgr~A* and improved dynamical imaging reconstructions, orbital optimization remains a target of investigation for LEO space-VLBI. Genetic or gradient searches for single-orbiter geometric improvements in $(u,v)$ coverage are a natural next step, while further identification of the constraints of realistic space launch will also reduce the space of possible orbits. These alternative paradigms for space expansions working in tandem with a LEO expansion are promising ways to improve angular resolution and will likely provide incentives for including different LEO orbits. Ultimately, future EHT results will inform what $(u,v)$-filling paradigms best serve the next generation of  science goals of high frequency VLBI.

\acknowledgements{We thank Ramesh Narayan for suggesting the $(u,v)$ filling fraction diagnostic, as well as Maura Shea and Vincent Fish for their discussions of space expansions to the EHT. We are grateful to Joseph Lazio for facilitating early discussions of the feasibility of high frequency VLBI from space. Jonathan Weintroub, Alex Raymond and Kari Hayworth were instrumental in understanding space-VLBI hardware requirements. We thank Freek Roelofs and Heino Falcke for their discussion of the science possibilities of space-VLBI, for their hospitality during the Future of High-Resolution Radio Interferometry in Space workshop, and for fostering collaboration around space-VLBI within the radio astronomy community at large. We are grateful to Avery Broderick for providing simulations of the orbiting ``hot spot.'' Finally, we thank our referee for their careful and thoughtful feedback on our manuscript. We thank the National Science Foundation (AST-1440254, AST-1716536) and the Gordon and Betty Moore Foundation (GBMF-5278) for financial support of this work. This work was supported in part by the Black Hole Initiative at Harvard University, which is supported by a grant from the John Templeton Foundation.}

\bibliographystyle{yahapj}
\bibliography{references}

\end{document}